\documentclass[paper]{JHEP}
\usepackage{amsmath}
\usepackage{epsfig}

\def\al{\alpha}
\def\as{\alpha_{\mbox{\scriptsize s}}}
\def\aef{\alpha_{\mbox{\scriptsize eff}}}

\def\qq{q\bar{q}}
\def\ee{e^+e^-}

\def\MSbar{\overline{\mbox{MS}}}
\def\MSbar{\overline{\mbox{\scriptsize MS}}}
\def\GeV{\mathop{\rm Ge\!V}}
\def\hh{hh}

\def\be{\beta}

\def\sig{\hat{\sigma}}
\def\eps{\epsilon}
\def\de{\delta}

\def\om{\omega}
\def\gam{\gamma}

\def\sh{\hat{s}}

\def\out{\mbox{\scriptsize out}}
\def\mat{\mbox{\scriptsize mat}}

\def\exact{\mbox{\scriptsize exact}}
\def\remnant{\mbox{\scriptsize remnant}}
\def\cO#1{{\cal{O}}\left(#1\right)}
\def\half{\mbox{\small $\frac{1}{2}$}}

\def\VEV#1{\left\langle#1\right\rangle}

\def\PT{\mbox{\scriptsize PT}}
\def\NP{\mbox{\scriptsize NP}}

\def\conf{\delta}
\def\cp{\lambda^{\NP}}

\def\bnu{\bar{\nu}}
\def\Ko{K_{\out}}
\def\bKo{\bar{K}_{\out}}
\def\tKo{{\tilde K}_{\out}}
\def\ka{\kappa}
\def\vka{\vec{\ka}}                

\def\cF{{\cal{F}}}

\def\cM{{\cal{M}}}
\def\cN{{\cal{N}}}
\def\cR{{\cal{R}}}
\def\cP{{\cal {P}}}
\def\cI{{\cal {I}}}
\def\cC{{\cal {C}}}
\def\cS{{\cal {S}}}
\def\cA{{\cal{A}}}
 \newskip\humongous \humongous=0pt plus 1000pt minus 1000pt
   \newif\ifdtup

\def\fun#1#2{\lower3.6pt\vbox{\baselineskip0pt\lineskip.9pt
  \ialign{$\mathsurround=0pt#1\hfil##\hfil$\crcr#2\crcr\sim\crcr}}}

\title{Out-of-plane QCD  radiation 
\\in hadronic \boldmath{$Z_0$} production\thanks{This research was
    supported in part by the EU Fourth Framework Programme, `Training
    and Mobitily of Researchers', Network `Quantum Chromodynamics and
    the Deep Structure of Elementary Particles, contract
    FMRX-CT98-0194
    (DG12 - MIHT).}}

\author{
  A.~Banfi, G.~Marchesini, G.~Smye,\\
  Dipartimento di Fisica, Universit{\`a} di Milano--Bicocca and INFN,
  Sezione di Milano, Italy} \author{
  G.~Zanderighi.\\
  Dipartimento di Fisica Nucleare e Teorica, Universit{\`a} di Pavia
  and INFN, Sezione di Pavia, Italy}

\abstract{We present the QCD analysis of the cumulative
  out-of-event-plane momentum distribution in the process $p{\bar p}$
  into $Z_0$ and a hard jet (event plane defined by the $p{\bar p}$
  and $Z_0$ momenta). Particular attention is placed on the
  near-to-planar events for which we derive the all-order resummed
  result to next-to-leading accuracy. We consider also the leading
  power correction originating from the fact that, even in hard
  processes, the resummed QCD coupling runs into the infrared region.
  We aim at the same level of accuracy which, in $\ee$ annihilation,
  seems to be sufficient for making predictions.  Contributions from a
  ``soft underlying event'' due to beam remnant interactions are
  discussed.  Experimental data (not yet available) are needed to cast
  light on the predictive level of standard QCD analysis in hard
  hadron-hadron collisions. We plot examples of the predicted
  distribution at Tevatron energies.  The techniques here developed
  can be extended to other hard hadron-hadron and hadron-lepton
  processes.}

\preprint{
  Bicocca--FT--01/16\\
  Pavia--FNT/T-01/15\\
  hep-ph/0106278\\
  June 2001}
\begin{document}

\section{Introduction}
QCD radiation (jet shape distributions) in $\ee$ annihilation has been
intensively studied at the accuracy needed to make quantitative
predictions \cite{PTstandard}--\cite{broad} (double and single
logarithmic resummations, matching with fixed order exact results and
power suppressed corrections).  These distributions have provided
various tests of QCD \cite{Exp-shape} and measurements of the running
coupling \cite{Exp-running}.

For hard hadron-hadron (\hh) collisions, the analysis of the
associated QCD radiation is more complex than in $\ee$ annihilation.
There are three main differences between the two processes.  First,
the analyzed $\ee$ jet-shape distributions are collinear and infrared
safe (CIS) quantities (all soft and collinear divergences cancel).  On
the contrary, in hard \hh-collisions \cite{DDTetc}, jet-shape
distributions are finite only after factorizing collinear singular
contributions from initial state radiation (giving rise to incoming
parton distributions at the appropriate hard scale).

A second important difference between the two processes is that, while
in hard hh-collisions a typical event is a multi-jet
emission\footnote{ Actually in the particular case of processes in
  which only leptons are emitted in the elementary collision, one has
  just $2$ jets, originating from the incoming parton radiation.}
(originating from two incoming partons and from partons going out from
the hard elementary collision), in $\ee$ annihilation the bulk of
events is given by $2$-jet emission (originating from the primary
$\qq$ pair). Only recently have $3$-jet event shape $\ee$ observables
been studied: the thrust minor $T_m$ \cite{Tmin} and the $D$-parameter
\cite{Dpar} in the near to planar $3$-jet limit and the light-jet mass
and narrow-jet broadening \cite{BG,DS} in the $2$-jet limit. In this
paper we are mainly interested in the near to planar $3$-jet limit.

Finally, in hard hh-collisions the standard QCD description does not
account for the entire emitted radiation.  One needs to add a soft
underlying event, not present in $\ee$ annihilation, which could be
considered to result from low-$p_t$ interactions involving the
spectator partons (beam remnant interaction). The necessity of adding
such a component to the radiation was studied in \cite{MW}, in which
was analysed the ``pedestal height'' in hard jet production, i.e.  the
mean transverse energy per unit rapidity accompanying a
high-transverse-energy jet.  The pedestal height and its jet
transverse energy dependence measured at the CERN $p\bar p$ collider
\cite{U1} are accounted for by superimposing on the standard hard QCD
emission a soft underlying event similar to that of a minimum-bias
collision.  For more recent analysis on the features of the beam
remnant interaction see \cite{MCworkshops}.

In this paper we consider the process
\begin{equation}
  \label{eq:process}
h_1+h_2\to Z_0\>+\>\mbox{jet} + \ldots
\end{equation}
with $h_1,h_2$ the two incoming hadrons (where $h_1h_2$ are $pp$ or
$p\bar p$). The weak $Z_0$ boson is emitted with large transverse
momentum $Q_t$.  Here the dots represent initial state jets and
intra-jet hadrons.  Defining the event plane as the plane formed by
the $h_1,h_2$ and $Z_0$ momenta, we consider the distribution in the
observable $\Ko$ defined as
\begin{equation}
  \label{eq:Kout}
  \Ko={\sum_h}'\,|{p_{h}^{\out}}|\>,
\end{equation}
where $p_{h}^{\out}$ is the out-of-plane momentum of the hadron $h$.
To avoid measurements in the beam regions, the sum indicated by
$\sum_h'$ extends over all hadrons not in the beam direction, i.e.
emitted outside a cone around the beams.  

The process \eqref{eq:process} involves three jets: the large angle
jet (generated by the hard parton recoiling against the weak boson)
and the two initial state jets (generated by the incoming partons).
The observable $\Ko$ is similar to $T_m$, the $3$-jet shape observable
in $\ee$ annihilation, in which the event plane is defined by the
thrust and the thrust major axes. Our analysis of $\Ko$ will then make
use of the methods introduced for the study of $T_m$ \cite{Tmin}.  We
will obtain the following factorized perturbative (PT) contributions
and $1/Q_t$ power corrections:
\begin{itemize}
\item incoming parton distributions at the proper hard scale. They
  will be obtained from resummations of all powers $\as^n\ln^n\mu/\Ko$
  ($\mu$ is the small factorization scale needed to subtract the
  collinear singularities and, as we shall see, $\Ko$ is the hard
  scale for these distributions);
\item ``radiation factors'' characteristic of our observable.  They
  will be obtained, for small $\Ko$, from resummations of double
  logarithmic (DL) and single logarithmic (SL) terms ($\as^n\ln^{n+1}
  \Ko/Q_t$ and $\as^n\ln^{n} \Ko/Q_t$ respectively).  
  From the collinear contributions we need to exclude the pieces
  already accounted for by the reconstruction of the hard scale $\Ko$
  in incoming parton distributions. 
  As we shall see, the hard scales in the radiation factor, of order
  $Q_t$, are determined by the geometry of the hard elementary
  process. To identify the scales to this accuracy we need to work at
  the SL level;
\item matching of the above resummed result with the fixed order exact
  calculations. This allows us to obtain a description of the
  distribution in the full range of $\Ko$ (small or of order $Q_t$);
\item leading $1/Q_t$--power corrections to the PT result for the
  radiation factor. The incoming parton factor has corrections which
  are of second order ($1/Q_t^2$), see \cite{DMW}, and they will not
  be considered here. 
\end{itemize}
The first three points belong to standard PT analysis.  Concerning the
last point, these non-perturbative (NP) corrections originate from the
fact that, resumming the PT expansion, one reconstructs the running
coupling $\as(k_t)$ at the virtual scale which assumes values between
$k_t\sim Q_t$ and $k_t=0$. They should not be confused with soft
underlying event contributions due to beam remnant interaction, which
will be considered later.

To deal with the running coupling in the small $k_t$ region we use the
same procedure followed in the analysis of $\ee$ jet-shape
distributions. We extrapolate the running coupling into the large
distance region using the dispersive approach \cite{DMW} and we
determine the coefficient of $1/Q_t$ corrections in terms of a single
parameter, usually denoted by $\al_0$, which is given by the integral
of the QCD coupling over the region of small momenta $k\le \mu_I$ (the
infrared scale $\mu_I$ is conventionally chosen to be $\mu_I=2\GeV$,
but the results are independent of its specific value).  Effects of
the non-inclusiveness of $\Ko$ will be included by taking into account
the Milan factor $\cM$ introduced in \cite{Milan} and analytically
computed in \cite{Milan2}.  The NP parameter $\al_0$, which is the
same for all jet shape observables linear in the transverse momentum
of emitted hadrons, has been measured and appears to be universal with
a reasonable accuracy~\cite{Exp-shape}.

There are a number of differences between the analysis of $\Ko$ in the
\hh-process \eqref{eq:process} and that of $T_m$ in $\ee$
annihilation, besides that concerning initial state radiation
mentioned above.  First of all in the present case the event plane is
defined by the $h_1,h_2,Z_0$ momenta while in $\ee$ it is defined by
the full structure of the emitted radiation. Then in $\ee$ one has a
complicated interplay between the observable and the event plane
definition (technically one has to introduce various Fourier
integration variables needed to factorize the event plane condition).
These complications are absent in the present case. A second important
difference is the presence of the recoil momenta of partons underlying
the three jets. The recoil components enter the observable, the
kinematics and the matrix element. In the $\ee$ case all three primary
partons generating the jets move out of the event plane, due to the
recoil with the emitted secondary partons. In the present case,
instead, the two incoming partons (at a low subtraction scale) are
fixed along the beam direction by the kinematics of the parton
process. This makes relevant the presence of the rapidity cut
excluding the beam region (see later).

The paper is organized as follows.  
In section \ref{sec:Observable} we define the process under
consideration, the observable $\Ko$ and its distribution. We specify
the phase space region of $\Ko$ in which we perform the QCD study.
In section \ref{sec:Parton} we introduce the factorized structure of
the distribution at parton level: 
the hard elementary partonic process, incoming parton distributions at
the hard scale and radiation factor corresponding to our observable.
In section \ref{sec:Factor} we perform the resummation at SL accuracy
(we start from the analysis of soft contributions). We show here how
to factorize the two contributions generating the incoming parton
distributions and the radiation factor.
In section \ref{sec:RadiationF} we obtain the PT contribution to the
radiation factor and its NP correction.
In section \ref{sec:Distribution} we compute the distribution, matched to
the exact fixed-order result, and we present some numerical results.
Finally, section \ref{sec:Discussion} contains a summary, discussions
and conclusions.  
We add few technical Appendices \ref{App:Born}-\ref{App:matching}. In
the last one (Appendix \ref{App:beam}) we discuss the contribution of
beam remnant interaction.

\section{The process and the observable \label{sec:Observable}}
The incoming hadrons and $Z_0$ momenta in the process
\eqref{eq:process} are given by
\begin{equation}
  \label{eq:hhq}
h_1=\half\sqrt{s}\,(1,0,0,1)\>,\quad 
h_2=\half\sqrt{s}\,(1,0,0,-1)\>,\quad
Q=(Q_0,0,Q_t,Q_z)\>,
\end{equation}
with large $Q_t$ of order of the $Z_0$ mass $M$. The $Z_0$ is taken
on-shell, and its decay products are excluded from our calculation.
Including the hadronic decays is possible by following the analysis of
$2$-jet emission in $\ee$ annihilation.  The observable $\Ko$ is given
by \eqref{eq:Kout} where $p_h^{\out}=p_{hx}$, since the event plane is
the $yz$-plane, and the sum extends over all hadrons with rapidity
$\eta_h$ in the range
\begin{equation}
\label{eq:rapidity}
|\eta_h|<\eta_0 \simeq
-\ln\tan\frac{\Theta_0}{2}\>,
\end{equation}
with $\eta_0$ large.  This implements a cut of angle $\Theta_0$ around
the two beam directions. To avoid a strong dependence on $\eta_0$ we
will consider $\Ko$ not too small (see later).

We study the integrated distribution in $\Ko$ at fixed $Q_t$:
\begin{equation}
  \label{eq:dsigma}
\begin{split}
  \frac{d\sigma(\Ko)}{d Q_t}=
\sum _{m}\int \frac{d\sigma_m} {d Q_t}\>
\Theta\left(\Ko-\sum_{h=1}^m{}' |p_{hx}|\right)\>,
\end{split}
\end{equation}
with $d\sigma_m/d Q_t$ the differential distribution for $m$ emitted
hadrons in the process under consideration. We then use this to analyse
the normalized $\Ko$ distribution for events with the $Z_0$ 
transverse momentum greater than some cut-off $Q_m$:
\begin{equation}
  \label{eq:Sigma}
\begin{split}
\Sigma(Q_m,\Ko)\!=\!\sigma^{-1}(Q_m)\int_{Q_m}^{Q_M}\!dQ_t
  \frac{d\sigma(\Ko)}{d Q_t}\,,\quad
\sigma(Q_m)\!=\!\int_{Q_m}^{Q_M}\!dQ_t
\sum _{m}\int \frac{d\sigma_m} {d Q_t}\>,
\end{split}
\end{equation}
with $Q_M$ a fixed upper limit. We will choose $Q_M$ at the
kinematical boundary.

\section{Parton process \label{sec:Parton}}
At parton level, the process \eqref{eq:process} is described by two
incoming partons of momenta $p_1,p_2$ (inside the hadrons $h_1,h_2$),
the outgoing $Z_0$ and an outgoing hard parton $p_3$ accompanied by an
ensemble of secondary partons $k_i$
\begin{equation}
  \label{eq:partonproc}
  p_1\,p_2 \to Q\,p_3\,k_1\cdots k_n\>.
\end{equation}
There are three configurations of the incoming partons, with $p_1,p_2$
corresponding to $\qq$, $qg$ and $gq$. 
Taking a small subtraction scale $\mu$ (smaller than any other scale
in the problem), we assume that $p_1,p_2$ (and the spectators) are
parallel to the incoming hadrons,
\begin{equation}
  \label{eq:pa}
p_1=\half\sqrt{s}(x_1,0,0,x_1)\>,\quad  p_2=\half\sqrt{s}(x_2,0,0,-x_2)\>.
\end{equation}
Therefore, the observable we study is
\begin{equation}
  \label{eq:Ko}
  \Ko=|p_{3x}|+{\sum_i}' |k_{ix}|\>.
\end{equation}
The hard parton $p_3$, recoiling against the weak boson, is emitted at
a large angle and near the event plane. For small $\Ko$ the secondary
parton momenta $k_i$ are also near the event plane.

The QCD calculation of the distribution \eqref{eq:dsigma} is based on
the factorization of parton processes:
\begin{itemize}
\item incoming parton distributions at the scale $\mu$;
\item elementary hard process;
\item evolution of the incoming parton distributions from $\mu$ to the
  hard scale $\Ko$ obtained by resumming contributions of partons collinear
  to $p_1$ and $p_2$;
\item radiation factor, which for small $\Ko$ is obtained by resumming
  contributions of partons soft and/or collinear to the three hard
  partons;
\item soft underlying event due to beam remnant interaction.
\end{itemize}
Concerning the last point, one expects a contribution to $\Ko$ from the
beam remnant interaction \cite{MW} which could be estimated by
\begin{equation}
  \label{eq:Krem}
\Ko^{\remnant}\simeq 2\eta_0\cN\,\VEV{|k_x|}^{\remnant}=
\frac{4\eta_0\,\cN}{\pi}\VEV{k_t}^{\remnant}\>,  
\end{equation}
with $\cN$ and $\VEV{k_t}^{\remnant}$ the mean number per unit
rapidity and the mean $k_t$ of hadrons produced in the beam remnant
interaction (see Appendix \ref{App:beam}). From the study in
\cite{MW,MCworkshops,MW88} one estimates $\Ko^{\remnant}$ of the order
of few GeV.  The models considered for the beam remnant interaction
(in the central rapidity region) do not depend on hard scales and
should therefore be the same in all hard processes.

In the next sections we discuss only the hard QCD pieces.  

\subsection{Incoming parton distributions at the scale $\mu$}
We denote by $q_a^{f}(x),\bar q_a^{f}(x)$ and $g_a(x)$, the
distributions in the momentum fraction $x$ of quark, antiquark and
gluon inside the hadron $h_a$ (with $a=1,2$) at the subtraction scale
$\mu$. The quark and antiquark carry the flavour index $f$.  We
introduce the incoming parton distributions at the scale $\mu$ for the
three configurations
\begin{equation}
  \label{eq:cP}
 \begin{split}
&\cP^{f}_{\qq}(x_1,x_2,\mu)\>
=\> q_{1}^{f}(x_1)\bar{q}_2^{f}(x_2)+ 
\bar{q}_1^{f}(x_1)q_2^{f}(x_2)\>,\\
&\cP^{f}_{qg}(x_1,x_2,\mu)\>
=\>\left[\,q_{1}^{f}(x_1)+\bar{q}_1^{f}(x_1)\,\right] g_2(x_2)\>,\\
&\cP^{f}_{gq}(x_1,x_2,\mu)\>
=\> g_1(x_1)\left[\,q_{2}^{f}(x_2)+\bar{q}_2^{f}(x_2)\,\right].
\end{split}
\end{equation}

\subsection{Elementary hard process}
We neglect the secondary emitted partons $k_i$ and the hard process
\eqref{eq:partonproc} reduces to 
\begin{equation}
  \label{eq:Eproc}
  P_1\,P_2\to Q\,P_3\>,
\end{equation}
where the incoming partons, the weak boson and the outgoing parton
momenta can be written as
\begin{equation}
  \label{eq:Ekin}
\begin{split}
&P_1=\half\sqrt{\sh}e^y(1,0,0,1)\>,\qquad
 P_2=\half\sqrt{\sh}e^{-y}(1,0,0,-1)\>,\\
&Q= \left(\>E\cosh y\!+\!p\cos\theta\sinh y,\>0,\>\>p\sin\theta,\>
    \>\>p\cos\theta\cosh y\!+\!E\sinh y\>\right),\\
&P_3=\left(\>p\cosh y\!-\!p\cos\theta\sinh y,\>0,-p\sin\theta,\>
    -p\cos\theta\cosh y\!+\!p\sinh y\>\right),\\
&p=\frac{\sh -M^2}{2\sqrt{\sh}}\>,\quad 
E=\frac{\sh +M^2}{2\sqrt{\sh}}\>,\quad
Q_t=p\sin\theta\>,\quad
\sh=X_1X_2s\>, \quad 
y=\half \ln \frac{X_1}{X_2}\>.
\end{split}
\end{equation}
Here $E,p$ and $\theta$ are the energy, momentum and angle of $Q$ with
$P_1$ in the centre of mass system of process \eqref{eq:Eproc}; $y$ is
the boost along the $z$-axis bringing the process \eqref{eq:Eproc}
from the centre of mass to the laboratory system; $X_1,X_2$ are the
momentum fractions of $P_1,P_2$ entering the elementary collision.

The elementary cross sections
${d\hat\sigma^{f}_{\conf}(\sh,Q_t)}/{dQ_t}$ for the three partonic
configurations are given in Appendix~\ref{App:Born}.  In the
elementary process \eqref{eq:Eproc} the outgoing momentum $P_3$ is in
the event plane so $\Ko=0$. Summing over flavours $f$ and incoming
parton configurations $\conf=\qq,qg,gq$, we have the Born contribution
to \eqref{eq:dsigma} as
\begin{equation}
\label{eq:dsigmaB}
  \frac{d\sigma_{0}}{d Q_t}=\int dX_1dX_2
\sum_{f,\,\conf}\left\{\frac{d\hat \sigma^{f}_{\conf}(\sh,Q_t)}
{dQ_t}\cdot \cP^{f}_{\conf}(X_1,X_2,\mu)\right\}.%,\qquad \sh=X_1X_2s\>.
\end{equation}

\subsection{Distribution at parton level} 
Considering the secondary emitted partons $k_i$, the hard momentum
$P_3$ moves out of the event plane, acquires a soft recoil $q_3$ and
the observable is
\begin{equation}
  \label{eq:partonKo}
\Ko={\sum_i}' |k_{ix}|+|q_{3x}|\>,
\qquad p_3=P_3+q_3\>, \qquad q_{3x}=-\sum_i k_{ix}\>.
\end{equation}
For $\Ko\sim Q_t$ the distribution is obtained from exact fixed order
results.  In the following we consider the region of small $\Ko$ in
which one needs all order QCD resummations.  In this region the
distribution \eqref{eq:dsigma} can be factorized as follows
\begin{equation}
  \label{eq:dsigma-fact}
\frac{d\sigma(\Ko)}{d Q_t} = \int_{\sh_0}^{s}\frac{d\sh}{s} 
\sum_{f,\,\conf} \left\{C_{\conf}(\as)\cdot
\frac{d\hat \sigma^{f}_{\conf}(\sh,Q_t)}{dQ_t}\cdot
\cI^{f}_{\conf}\left(\sh,Q_t,\Ko\right)\right\},
\end{equation}
with the lower bound $\sh_0$ given in \eqref{eq:sh0}. The distribution
$\cI^{f}_{\conf}$ includes the incoming parton distribution $\cP(\mu)$
in \eqref{eq:cP} and resums the (factorized) collinear and soft powers
of $\ln\mu/\Ko$ or $\ln\Ko/Q_t$, with $\mu$ the subtraction scale for
collinear singularities.

In \eqref{eq:dsigma-fact} we have factored out the elementary parton
cross sections $d\hat\sigma/dQ_t$ given in \eqref{eq:dQt2}.
The coefficient $C_{\conf}(\as)=1+\cO{\as}$ is a non-logarithmic
function which takes into account hard corrections not included in
$\cI^{f}_{\conf}$. It can be computed from the exact fixed order
results.
As one expects, and as will be discussed in detail in the next
section, the distribution $\cI^{f}_{\conf}$ can be factorized as
follows
\begin{equation}
  \label{eq:Factor}
  \cI^{f}_{\conf}(\sh,Q_t,\Ko)=
\int dX_1dX_2\,\de\left(\frac{\sh}{s}-X_1X_2\right)\>     
\cP^{f}_{\conf}\left(X_1X_2,\Ko\right)\cdot
      \cA_{\conf}\left(\sh,y,Q_t,\Ko\right),
\end{equation}
with $y=\half\ln X_1/X_2$. We have two factors:
\begin{itemize}
\item the first, $\cP^f_\conf$, is the incoming parton probability
  evolved from $\mu$ to the hard scale $\Ko$. It resums singular terms
  coming from secondary partons which are collinear to the incoming
  partons $p_1$ and $p_2$, giving rise to the anomalous dimensions;
\item the second, $\cA_{\conf}$, is the radiation factor corresponding
  to the observable $\Ko$.  It resums powers of $\ln \Ko/Q_t$ and is a
  CIS quantity.  It is sensitive only to QCD radiation and therefore
  does not depend on the flavour (we neglect quark masses).  There are
  various hard scales in $\cA_{\conf}$ (given in terms of the
  invariants $(P_a P_b)\sim Q_t^2$) which will be determined by the SL
  accuracy analysis.
\end{itemize}
In the following we will obtain $\cI^{f}_{\conf}$ by resumming the QCD
radiation in the rapidity region \eqref{eq:rapidity}. To simplify the
analysis we will consider $\Ko$ sufficient large, in the region
\begin{equation}
  \label{eq:minKo}
  \Ko> \Ko^c\sim M\,e^{-\eta_0}\>.
\end{equation}
As we will discuss in detail, here the PT results at SL level do not 
depend on $\eta_0$, (hadrons $h'$ emitted inside the beam cones
typically have $|p_{h'x}|$ smaller than $\Ko^c$).  The dependence on
$\eta_0$ and on the boost $y$ enters only in the NP corrections of
$\cA_{\conf}$ (NP corrections affect the distribution at any
rapidity).

Ideally, in order for our results to be valid over the widest possible
range of $\Ko$, we would like to take $\eta_0$ as large as is
experimentally possible.

\section{Resummation and factorization of $\cI^{f}_{\conf}$ 
\label{sec:Factor}} 
In this section we derive the factorization structure
\eqref{eq:Factor}.  In particular we show that, to our accuracy, in
$\cP_{\conf}^f$ the hard scale is actually given by $\Ko$ and the
rapidity cut $\eta_0$ is irrelevant. We also deduce the expression of
the radiation factor $\cA_{\conf}$ to next-to-leading order which
will be discussed in the next sections.

We consider first the resummation of logarithmic terms coming from
soft secondary partons. They include all DL terms and the SL terms
originating from soft partons.  The remaining SL contributions
(collinear non-soft secondary partons) will be included later. They
give the non-soft part of the anomalous dimension and contribute to
fix the hard scales in $\cA_{\conf}$, see \cite{Tmin}.

We resum the enhanced soft contributions to next-to-leading order by
using the factorization of soft radiation.  To this end we extend to
the process \eqref{eq:partonproc} the methods previously introduced in
$\ee$ to analyse the distribution of CIS observables for $3$-jet
events, see \cite{Tmin}.  The new fact in the present case is that the
distribution is not a CIS observable and depends on the subtraction
scale $\mu$.

The square amplitude for the emission of $n$ soft partons in the
process \eqref{eq:partonproc} can be factorized as follows 
\begin{equation}
  \label{eq:Mn1}
  |M^{f}_{\conf,\,n}(k_1\ldots k_n)|^2
\simeq|M^{f}_{\conf,\,0}|^2\cdot S_{\conf,\,n}(k_1\ldots k_n)\>.
\end{equation}
The first factor is the Born square amplitude which gives rise to the
Born distribution $d\hat \sigma^{f}_{\conf}/{dQ_t}$ in
\eqref{eq:dsigma-fact}. The second factor is the distribution in the
soft partons emitted from the system of the three hard partons
$p_1,p_2$ and $p_3$.  It depends on the colour charges of the
emitters in the various configurations $\conf$.

By using \eqref{eq:Mn1}, the soft contributions to the distribution
$\cI^{f}_{\conf}$ are resummed by
\begin{equation}
  \label{eq:cIsoft}
\begin{split}
  &\cI^{f}_{\conf}(\sh,Q_t,\Ko)=\int dx_1dx_2\,
  \cP^{f}_{\conf}(x_1,x_2,\mu) \>\sum_n\,\frac{1}{n!}\int
  \prod_{i=1}^n \>\frac{d^3k_i}{\pi\om_i}\>S_{\conf,\,n}(k_1\ldots
  k_n) \\& \times 
  \delta\left(\frac{\sh}{s}-X_1X_2\right)\>
dq_{3x}\,\delta(q_{3x}+\sum_i k_{ix})\>  
\Theta\left(\Ko-{\sum_i}'|k_{ix}|-|q_{3x}|\right).
\end{split}
\end{equation}
Here the momentum fractions $X_1,X_2$ of the hard elementary process
\eqref{eq:Eproc} are given by
\begin{equation}
  \label{eq:Xa}
  X_a=x_a\prod_{i\in \cC_a}z_i\>, \qquad a=1,2,
\end{equation}
where $z_i$ is the splitting fraction associated with collinear
radiation from the incoming partons, and $\cC_{a}$ is the region in
which $k_i$ is collinear to $p_a$.  As we shall discuss later, the
precise form of the collinear regions is not crucial.  The soft factor
$S_{\conf,\,n}$ depends on the hard elementary collision variables
$\sh,y,Q_t$ in \eqref{eq:Ekin} and the recoil momentum $q_{3x}$.

To resum the expansion we write the constraint on $\sh$ in the form
\begin{equation}
  \label{eq:MF1}
  \de\left(\frac{\sh}{s}-x_1x_2\!\prod_{\cC_1,\cC_2}z_i\right)=
\prod_{a=1}^{2}\int_0^1\frac{dX_a}{X_a}\!\int\frac{dN_a}{2\pi i}
\left(\frac{x_a}{X_a}\right)^{N_a-1} \prod_{i\in \cC_{a}}z_i^{N_a-1}\cdot
\de\left(\frac{\sh}{s}-X_1X_2\right),
\end{equation}
and the phase space of \eqref{eq:cIsoft} in terms of Fourier and Mellin
transforms
\begin{equation}
  \label{eq:MF2}
\begin{split}
&\Theta\left(\Ko-{\sum_i}'|k_{ix}|-|q_{3x}|\right)
\delta(q_{3x}+\sum_i k_{ix}) \\
&\quad=\int\frac{d\nu\,e^{\nu\Ko}}{2\pi i\nu}
\int_{-\infty}^{\infty}\frac{\nu d\be}{2\pi}\,e^{-\nu(|q_{3x}|-i\be q_{3x})}
{\prod_i}' e^{-\nu|k_{ix}|}\>\prod_i e^{i\nu\be k_{ix}}\>.
\end{split}
\end{equation}
The Mellin $N_a$- and $\nu$-contours run parallel to the imaginary
axis with Re$\,N_a\!>\!1$ and Re$\>\nu\!>\!0$.  Using \eqref{eq:MF1}
and \eqref{eq:MF2}, $\cI(\Ko)$ can be written in the form
\begin{equation}
\label{eq:Sigma1}
\begin{split}
\cI^{f}_{\conf}(\sh,Q_t,\Ko)=
&\int\frac{d\nu\,e^{\nu\Ko}}{2\pi i\nu}\!\!
\prod_{a=1}^{2}\int_0^1dX_a \int_0^1 \frac{dx_a}{X_a}\>\>
\int\frac{dN_a}{2\pi i} \left(\frac{x_a}{X_a}\right)^{N_a-1}
\\&\times
\de\left(\frac{\sh}{s}-X_1X_2\right)
\cP^{f}_{\conf}(x_1,x_2,\mu)
\cdot \cS_{\conf}(\nu,N_1,N_2,\mu)\>,
\end{split}
\end{equation}
where the distribution $\cS_{\conf}(\nu,N_1,N_2,\mu)$ is obtained by
resummming the soft contributions 
\begin{equation}
  \label{eq:cS}
\begin{split}
&\cS_{\conf}(\nu,N_1,N_2,\mu)=\int_{-\infty}^{\infty}
\frac{\nu d\be dq_{3x}}{2\pi}\>e^{-\nu(|q_{3x}|-i\be q_{3x})}
\>e^{-\cR_{\conf}(\nu,\be,N_1,N_2,\mu)}\>,\\
&e^{-\cR_{\conf}(\nu,\be,N_1,N_2,\mu)}=
\>\sum_n\,\frac{1}{n!}\int\prod_{i=1}^n \> 
\left\{\frac{d^3k_i}{\pi\om_i}\, U(k_{i})\>\eps_{12}(z_i)\right\}
\>S_{\conf,\,n}(k_1\ldots k_n)\>.
\end{split}
\end{equation}
Here the source $U(k)$ takes into account the phase space
constraints ($e^{i\nu\be k_x}$) and the observable ($e^{-\nu|k_x|}$) 
for rapidity $\eta_k$ in the region \eqref{eq:rapidity}
\begin{equation}
  \label{eq:U}
\left\{\>\>
\begin{split}
&U(k)= u(k_x)\:\,\equiv 
e^{-\nu(|k_{x}|-i\be k_{x})}\quad\, \mbox{for}\> |\eta_k|<\eta_0\>,\\
&U(k)= u_0(k_x)\equiv 
e^{i\nu\be k_{x}} \qquad \qquad \mbox{for}\> |\eta_k|>\eta_0\>.
\end{split}
\right.
\end{equation}
The source $\eps_{12}(z)$ takes into account the fact that the
collinear radiation reduces the longitudinal momentum components
of the incoming partons $p_1$ and $p_2$.
\begin{equation}
  \label{eq:eps12} 
\left\{\>\>
\begin{split}
&\eps_{12}(z)=z^{N_a-1}\>,
\quad \mbox{for}\> k \in \cC_{a}\>,\quad a=1,2\\
&\eps_{12}(z)=1\>,\>\>\>\>
\qquad \mbox{for}\> k \in\!\!\!\!|\>\> \cC_{1}, \cC_{2}\>.
\end{split}
\right.
\end{equation}
The near-to-planar region $\Ko \ll Q_t$  corresponds to the region
of the Mellin variable $\nu Q_t\gg 1$. Having introduced the Fourier
variable $\nu\be$ conjugate to $k_x$, the condition $|k_x|\sim\Ko\ll
Q_t$ corresponds to $\be=\cO{1}$ (in other words, the $\be$-integration
will be fastly convergent). Therefore we have here that $\nu Q_t$ is
the only large parameter we need to consider.

To obtain the exponent $\cR_{\conf}$ at SL accuracy we follow the same
procedure described in detail in \cite{Tmin} and for the configuration
$\conf$ we have
\begin{equation}
  \label{eq:cRad}
\cR_{\conf}(\nu,\be,N_1,N_2,\mu)=\int\frac{d^3k}{\pi\om}W_{\conf}(k)\,
\left[1-U(k)\eps_{12}(z)\right].
\end{equation}
Here $W_{\conf}(k)$ is the distribution of soft gluon radiation off
the hard three-parton antenna in the configuration ${\conf}$ given by
\begin{equation}
  \label{eq:W}
\begin{split}
&W_{\qq }(k)=\frac{N_c}{2}\left(w_{13}+w_{23}-\frac{1}{N_c^2}w_{12}\right),\\
&W_{qg }(k)=\frac{N_c}{2}\left(w_{12}+w_{23}-\frac{1}{N_c^2}w_{13}\right),\\
&W_{gq }(k)=\frac{N_c}{2}\left(w_{12}+w_{13}-\frac{1}{N_c^2}w_{23}\right),
\end{split}
\end{equation}
where $w_{ab}(k)$ is the standard soft distribution for the emission
of a soft gluon $k$ from the $ab$-dipole 
\begin{equation}
  \label{eq:wab}
w_{ab}(k)\>=\> \frac{\as(k_{ab,\,t})} {\pi\,k^2_{ab,\,t}} \>,
\qquad k^2_{ab,\,t}= \frac{2(p_ak)(kp_b)}{(p_ap_b)}\>.
\end{equation}
Here the running coupling is taken in the physical scheme \cite{CMW}
and $k_{ab,\,t}$ is the invariant transverse momentum of $k$ with
respect to the $p_a,p_b$ hard partons.  The unity in the square
bracket in \eqref{eq:cRad} takes into account the virtual corrections.
The expression in \eqref{eq:cRad} resums all enhanced terms at
next-to-leading order coming from soft contributions.  To reach the
full SL accuracy, one needs to take into account also the non-soft
part of the collinear splitting which will be considered later.

The exponent $\cR_{\conf}$ in \eqref{eq:cRad} differs from the $\ee$
radiator in $3$-jet events by the presence of the $z^{N_a-1}$ factor
in the sources $\eps_{12}$ when $k$ is collinear to one of the
incoming partons $p_1$ or $p_2$. As a consequence, $\cR_{\conf}$ is
not a CIS quantity and then depends on the subtraction scale $\mu$.

\subsection{Factorizing incoming parton distributions and radiation factor}
In the present formulation, the factorization structure
\eqref{eq:Factor} results by splitting the source as follows
\begin{equation}
  \label{eq:split}
  [1-U(k)\eps_{12}(z)]=[1-U(k)]\>+\>[1-\eps_{12}(z)]U(k)\>,
\end{equation}
so that the exponent $\cR$ can be split into two terms 
\begin{equation}
  \label{eq:R+G}
   \cR_{\conf}(\nu,\be,N_1,N_2,\mu)= R_{\conf}(\nu,\be)+
\Gamma_{\conf}(\nu,\be,N_1,N_2,\mu)\>.
\end{equation}

The first term, which produces the radiation factor in
\eqref{eq:Factor} is given by
\begin{equation}
\label{eq:Rad}
R_{\conf}(\nu,\be)=\int\frac{d^3k}{\pi\om}\>W_{\conf}(k)\>[1-U(k)]\>.
\end{equation}
It is a CIS quantity, independent of $N_a$, of the same type as the
radiator in $3$-jet $\ee$ processes \cite{Tmin,Dpar}.  It
depends also on the hard variables $\sh,y,Q_t$ in \eqref{eq:Ekin}, on
the rapidity cut $\eta_0$ and on the recoil component $q_{3x}$. 

The second term is given by
\begin{equation}
\label{eq:Gam}
\Gamma_{\conf}(\nu,\be,N_1,N_2,\mu)=
\sum_{a=1}^{2}\int_{\cC_{a}}\frac{d^3k}{\pi\om}\>
W_{\conf}(k)\> \left[1-z^{N_a-1}\right]\>U(k)\>.
\end{equation}
The integration is confined to $k$ within the region $\cC_a$ collinear
to $p_a$, so that we can neglect the dependence on $q_{3x}$, in the
soft limit. $\Gamma_{\conf}$ is collinear singular and we therefore
need to introduce here the subtraction scale $\mu$. This term gives
the (soft part of the) anomalous dimensions of the two incoming partons,
and so evolves the incoming parton distribution $\cP(\mu)$ to
$\cP(\Ko)$ in \eqref{eq:Factor}.

We discuss first $\Gamma_{\conf}$ and then  the radiator $R_{\conf}$.

\subsection{Incoming parton evolution at the hard scale \label{sec:Gamm}}
We first observe that, to our accuracy, $\Gamma_{\conf}$ does not
depend on the rapidity cut \eqref{eq:rapidity}. 
This is shown by the following argument. For large $\nu Q_t$ we have
$\nu\sim \nu|\be|\sim\Ko^{-1}$ so that the difference between the two
sources
\begin{equation}
  \label{eq:udiff}
u_0(k_x)-u(k_x)=e^{i\nu\be k_x}\>\left(1-e^{-\nu|k_x|}\right),
\end{equation}
vanishes unless $|k_x|\sim\Ko$. Therefore, in the integral giving
$\Gamma_{\conf}$, we can replace the source $U(k)$ with $u(k_x)$ (or
with $u_0(k_x)$) with correction of order $\as(\Ko)$. The result is
then independent of the rapidity cutoff $\eta_0$, to our accuracy.

The calculation of $\Gamma_{\conf}$ is rather standard and is
performed in Appendix \ref{App:Gamm}. The crucial point here is the
identification of the hard scale. In our case the scale is obtained 
by using the fact that, within next-to-leading order accuracy, the
$k_x$-source can be replaced by an effective cutoff (see for
instance Appendix~C of Ref.~\cite{Dpar})
\begin{equation}
  \label{eq:rho}
[1-u(k_x)]\>\Rightarrow\>
\Theta\left(\,|k_x|-\rho^{-1}\,\right)\>,\qquad 
\rho=e^{\gam_E}\nu\sqrt{1+\be^2}\>.
\end{equation}
From this one has that the hard scale is of order $\rho^{-1}$.
From Appendix \ref{App:Gamm} one has
\begin{equation}
  \label{eq:gam}
\begin{split}
\Gamma_{\conf}(\nu,\be,N_1,N_2,\mu)\simeq 
-\int_{\mu}^{\rho^{-1}}\!\frac{dk}{k}\,\gam_{\conf}(N_1,N_2,\as(k))\>,
\end{split}
\end{equation}
Here $\gam_{\conf}$ is the soft part of the anomalous
dimensions of the two incoming partons
\begin{equation}
  \label{eq:gamN}
\gam_{\conf}(N_1,N_2,\as(k))=-\frac{2\as(k)}{\pi}\>
\sum_{a=1}^{2}C_a^{(\conf)}\,S_1(N_a)\>,
\end{equation}
where the function $S_1(N)$ is given in \eqref{eq:S1} and
\begin{equation}
  \label{eq:Colour}
  C_1^{\qq}=C_1^{qg}=
  C_2^{\qq}=C_2^{gq}=
  C_3^{qg}=C_3^{gq}=C_F\>,\quad
  C_1^{gq}=C_2^{qg}=C_3^{\qq}=C_A\>,
\end{equation}
are the colour charges of the parton $\#a$ in the configuration
$\conf$.  The soft part of the anomalous dimension \eqref{eq:gamN} is
accurate at two loop order provided we use the coupling in the
physical scheme \cite{CMW}.  It is diagonal in the configuration index
since soft radiation is universal and does not change the nature of
incoming parton.

Upon integration over $x_a$ and the Mellin variables $N_a$ one obtains 
\begin{equation}
  \label{eq:cPhard}
\prod_{a=1}^{2}\int_0^1\frac{dx_a}{X_a}\int\frac{dN_a}{2\pi i} 
\left(\frac{x_a}{X_a}\right)^{N_a-1}\!
\cP_{\conf}^{f}(x_1,x_2,\mu)\cdot
e^{-\Gamma_{\conf}(\nu,\be,N_1,N_2,\mu)}
\,\simeq\, \cP_{\conf}^{f}(X_1,X_2,\Ko)\>.
\end{equation}
The $\mu$ dependence cancels in the product $\cP(\mu)\cdot
e^{-\Gamma(\mu)}$ giving the distribution $\cP(\rho^{-1})$ with
$\rho$ given in \eqref{eq:rho}.  For small $\Ko$ we have $\nu^{-1}\sim
\Ko$ and we can replace the hard scale $\rho^{-1}$ with $\Ko$
with corrections of order $\as(\Ko)$ not enhanced by logarithms.

Up to now we have considered only soft contributions giving
\eqref{eq:gamN}, the leading part of the anomalous dimension for large
$N_a$. However for the process under consideration we have to consider
contributions for $N_a =\cO{1}$ ($X_a$ not too close to one) and we
need to consider the full anomalous dimension which is no longer
diagonal in the configuration index $\conf$. In this way the incoming
partons may change from a quark to a gluon and vice versa.  The
resulting exponent $\Gamma$ becomes a matrix in $\conf$ so that the
product $\cP(\mu)\cdot e^{-\Gamma(\mu)}$ in \eqref{eq:cPhard} becomes
a matrix product and one obtains the parton distribution
$\cP_{\conf}^{f}(X_1,X_2,\Ko)$ fully evolved from the subtraction to
the hard scale. We do not consider here power corrections since they
are of second order ($1/Q_t^2$), see \cite{DMW}.

Finally, \eqref{eq:cPhard} gives the factorized result \eqref{eq:Factor}.

\section{Radiation factor \label{sec:RadiationF}}
The distribution $\cA_{\conf}(\sh,y,Q_t,\Ko)$ entering into the
factorized expression \eqref{eq:Factor} is obtained from the piece
$R_{\conf}(\nu,\be)$ of the full radiator in \eqref{eq:cRad}. From
\eqref{eq:cS} we obtain
\begin{equation}
\label{eq:Mellin}
\cA_{\conf}(\sh,y,Q_t,\Ko)\>=\>
\int\frac{d\nu\,e^{\nu\Ko}}{2\pi i\nu}\>
\hat \cA_{\conf}(\sh,y,Q_t,\nu)\>,
\end{equation}
where 
\begin{equation}
  \label{eq:hatsigma}
  \hat \cA_{\conf}(\sh,y,Q_t,\nu)=
\int_{-\infty}^{\infty}dq_{3x}\,
\int_{-\infty}^{\infty}\frac{\nu d\be}{2\pi}\,
e^{-\nu(|q_{3x}|-i\be q_{3x})}\>
e^{-R_{\conf}(\nu,\be)}\>.
\end{equation}
We recall that the CIS radiator $R_{\conf}(\nu,\be)$ depends on the
elementary collision variables $\sh,y,Q_t$ (see \eqref{eq:Ekin}),
rapidity cut $\eta_0$ and recoil component $q_{3x}$.  The radiator
$R_{\conf}$ contains a PT contribution and an NP correction
\begin{equation}
  \label{eq:Radtot}
  R_{\conf}(\nu,\be)=  R^{\PT}_{\conf}(\nu,\be)+
\de R_{\conf}(\nu)\>,  
\end{equation}
which we discuss in the following.

\subsection{The PT radiator \label{sec:RadPT}}
Here we obtain $R^{\PT}_{\conf}$ to SL accuracy.  We start from 
%the soft part which is given by 
\eqref{eq:Rad}. The contribution from the hard collinear splitting
will be considered later.

First of all, we can neglect the soft recoil $q_3$ since, as shown in
\cite{broad}, it contributes beyond SL accuracy.  Then we set $p_a\to
P_a$ (see \eqref{eq:Ekin}). Moreover, to SL accuracy, the PT radiator
is independent of $\eta_0$ and $y$, as long as one considers $\Ko$ in
the region \eqref{eq:minKo}. This is due to the fact that for
$|\eta_k|>\eta_0$ we can replace $u_0(k)$ with $u(k_x)$. The argument
is similar to the one presented for the evaluation of $\Gamma_{\conf}$
(see Appendix \ref{App:RadPT} for a detailed discussion on this
point).

The calculation of the PT radiator is performed in Appendix
\ref{App:RadPT} and, to SL accuracy, one finds
\begin{equation}
  \label{eq:Rad-PT}
  R_{\conf}^{\PT}(\rho)\>=
C_F\,r\left(\rho,Q^{\conf}_q\right)+
C_F\,r\left(\rho,Q^{\conf}_{\bar q}\right)+
C_A\,r\left(\rho,Q^{\conf}_g\right)\>.
\end{equation}
where 
\begin{equation}
  \label{eq:rab3}
r\left(\rho,Q_{a}\right)\equiv\>
\int_{\rho^{-1}}^{K} \frac{dk}{k}\frac{2\as(2k)}{\pi}\>
\ln\frac{Q_{a}}{2k}\>,
\qquad K\sim\sqrt{\sh}\>,
\end{equation}
with $\as$ in the physical scheme \cite{CMW}. The factor $2$ in the
argument of the running coupling takes into account the fact that the
observable involves only the $x$-component of transverse momentum,
while the $y$-component is integrated out. The upper limit $K$ in the
$k$ integration is beyond our accuracy as long as of order of the hard
scale $\sqrt{\sh}$. The PT hard scales are given by
\begin{equation}
  \label{eq:Qa}
\begin{split}
&Q^{\qq}_q=Q^{\qq}_{\bar q}=Q_{12}\cdot e^{-3/4}\>,\qquad 
Q^{\qq}_g=\frac{Q_{13}\,Q_{32}}{Q_{12}}\cdot e^{-\be_0/4N_c}\>,\\
&Q^{qg}_q=Q^{qg}_{\bar q}=Q_{13}\cdot e^{-3/4}\>,\qquad 
Q^{qg}_g=\frac{Q_{12}\,Q_{32}}{Q_{13}}\cdot e^{-\be_0/4N_c}\>,\\
&Q^{gq}_q=Q^{gq}_{\bar q}=Q_{23}\cdot e^{-3/4}\>,\qquad 
Q^{gq}_g=\frac{Q_{12}\,Q_{13}}{Q_{23}}\cdot e^{-\be_0/4N_c}\>.
\end{split}
\end{equation}
The first factors, expressed in terms of the $Q_{ab}$ in equation
\eqref{eq:Q_ab}, are determined by the large-angle soft emission.  The
quark or antiquark scale is given by the invariant mass of the
quark-antiquark dipole. The scale for the gluon is its transverse
momentum with respect to the quark-antiquark dipole.  The rescaling
constants $e^{-3/4}$ and $e^{-\be_0/4N_c}$ take into account SL
corrections coming from the hard parton splitting functions.  These
constants and the precise expression of the geometry dependent scales
$Q_a$ are important only beyond DL accuracy. In conclusion, the PT
radiator depends on the geometry of the event, on the underlying
configuration $\conf$, and on the presence of the recoil in the
kinematics ($\be$-dependence).  It does not depend on the boost $y$
since the rapidity cut $\eta_0$ does not affect the PT result.

\subsection{NP corrections to the radiator \label{sec:RadNP}}
The procedure for computing the leading NP corrections, including two
loop order to take into account the non-inclusiveness of jet
observables, is the usual one \cite{Milan} and recalled in Appendix
\ref{App:RadNP}. One finds
\begin{equation}
\label{eq:Rad-NP}
\de R_{\conf}(q_{3x})=\nu\,\cp
\left(C_{1}^{(\conf)}\,(\eta_0-\eta_3)+C_{2}^{(\conf)}\,(\eta_0+\eta_3)
+C_{3}^{(\conf)}\ln\frac{\zeta Q_t}{|q_{3x}|}\right),
\qquad \zeta=2e^{-2}\>,
\end{equation}
where $\cp$ is the NP parameter given in \eqref{eq:cp}. It is
expressed in terms of the integral of the running coupling over the
infrared region
\begin{equation}
  \label{eq:al0}
\al_0(\mu_I)=\int_0^{\mu_I}\frac{dk}{\mu_I}\,\as(k)\>.
\end{equation}
This parameter $\al_0(\mu_I)$ is the same as enters the $\ee$ jet
shape variables $1-T$, $C$, $M^2/Q^2$, $B$, $T_m$ and $D$, see
\cite{Tmin,Dpar,NPstandard}.  After merging PT and NP contributions to
the observable in a renormalon free manner, one has that the
distribution is independent of $\mu_I$.

The quantity $\eta_3$ is the rapidity of the outgoing hard parton
$\#3$ in the laboratory system \eqref{eq:Ekin}
\begin{equation}
\label{eq:eta3}
\eta_3 = \half\ln\frac{h_2P_3}{h_1P_3}=
y + \ln\frac{Q_{23}}{Q_{13}} = y+\ln\tan\frac{\theta}{2}\>.
\end{equation}
Due to the symmetry of the integrand of \eqref{eq:dsigma-fact} for
collisions of interest to us, we may take $\cos\theta>0$ as discussed
before \eqref{eq:dQt2}.  Finally, the colour charge $C_{a}^{(\conf)}$
of parton $\#a$ in the configuration $\conf$ is given in
\eqref{eq:Colour}.

The result has a simple interpretation based on the fact that the
observable is independent of the soft gluon rapidity.  First consider
the contribution from the outgoing parton $\#3$, proportional to
$C_{3}^{(\conf)}$. Here the $\ln |q_{3x}|$ contribution comes from the
rapidity integration along the outgoing parton direction.
Real-virtual cancellation, which takes place when the angle of the
outgoing parton $p_3$ with the event plane exceeds the corresponding
angle of the soft gluon, provides an effective rapidity cut and leads
to the $\ln |q_{3x}|$ term.
The rescaling factor $\zeta$ is due to the fact that rapidity is
related to the angle between two vectors while the boundary here is
given in terms of the angle between a vector and a plane.

Then consider the NP correction due to emission from the incoming
parton proportional to $C_{1}^{(\conf)}$ and $C_{2}^{(\conf)}$.  Since
the $p_1$ and $p_2$ momenta are fixed, no recoil is present and the
soft gluon rapidity $\eta_k$ is bounded by $\eta_0$. The
$\eta_k$-integration gives then $(\eta_0-\eta_3)$ and
$(\eta_0+\eta_3)$, for $a=1$ and $a=2$ respectively, i.e, the length
of the rapidity interval between $\eta_3$ and the boundary $\pm
\eta_0$. If we remove the $\eta_0$-bound (i.e. we keep in the
observable all partons including the ones in the beam direction) the
rapidity of the soft gluon can go up to the kinematical limit and one
obtains NP corrections involving the logarithmic moment of the running
coupling in the infrared region.

\section{Distribution \label{sec:Distribution}}
We are now in the position to obtain the full distribution
$\Sigma(Q_m,\Ko)$ in \eqref{eq:Sigma} to the standard QCD accuracy.
First we obtain the resummed PT expression $\Sigma^{\PT}$.  Then,
using the exact result of the $\cO{\as^2}$ matrix element calculation,
we compute the first correction of the coefficient function $C(\as)$
in \eqref{eq:dsigma-fact} and perform the matching of the resummed and
the exact result to this order.  We then include the leading power
correction coming from the NP part of the radiator \eqref{eq:Rad-NP}.
Finally we plot this distribution for $p\bar p$ collisions at the
Tevatron (the contribution of the underlying event due to the beam
remnant interaction can be taken into account successively as a ``rigid
shift'' in the argument by the quantity $\Ko^{\remnant}$ in
\eqref{eq:Krem}).

\subsection{Resummed PT contribution}
The PT contribution $\Sigma^{\PT}(Q_m,\Ko)$ to SL accuracy is obtained
from the radiation factor \eqref{eq:Mellin} by taking only the PT part
of the radiator given by \eqref{eq:Rad-PT}.  Performing the Mellin
transform (see Appendix \ref{App:Distribution}) we obtain, to SL
accuracy,
\begin{equation}
  \label{eq:sigPT}
  \cA^{\PT}_{\conf}(\sh,Q_t,\Ko)=
e^{-R_{\conf}^{\PT}\left(\Ko^{-1}\right)}\cdot
\frac{\cF(R')\,e^{-\gam_E R'}}{\Gamma\left(1+R'\right)}.
\end{equation}
Here $R_{\conf}^{\PT}\left(\Ko^{-1}\right)$ is given in
\eqref{eq:Rad-PT} with $\rho$ replaced by $\Ko^{-1}$ and
\begin{equation}
  \label{eq:cF}
R'=C_T\,\frac{2\as(\Ko)}{\pi}\ln \frac{K}{\Ko}\>,\qquad  
\cF(R')=\frac{\Gamma\left(\frac{1+R'}{2}\right)}
{\sqrt{\pi}\Gamma\left(1+\half R'\right)}\>, \quad C_T=2C_F+C_A\>,
\end{equation}
with $K$ the same hard scale as in \eqref{eq:rab3} so that
\begin{equation}
  \label{eq:sigPT=1}
    \cA^{\PT}_{\conf}(\sh,Q_t,K)=1\>.
\end{equation}
To first order in $\as$ we have
\begin{equation}
  \label{eq:sigPT1}
\cA^{\PT}_{\conf}(\sh,Q_t,\Ko)=
1-\frac{\as}{\pi}\!\left\{C_T L_K^2 \!+\!
2\sum_aC^{(\conf)}_a\ln\frac{Q_a^{(\conf)}}{K}\, L_K\right\}\!
+\cO{\as^2}, \>\> L_K\equiv\ln\frac{K}{\Ko} .
\end{equation}
The PT contribution to the (normalized and integrated) distribution in
\eqref{eq:Sigma} is given, to SL accuracy, by
\begin{equation}
\label{eq:SigPT-fine}
\begin{split}
\Sigma^{\PT}(Q_m,\Ko)=\frac{1}{\sigma(Q_m)}\int_{Q_m}^{Q_M} dQ_t 
&\int dX_1dX_2\sum_{f, \conf}
\left\{\frac{d\hat \sigma^{f}_{\conf}}{dQ_t}
\> \cP^{f}_{\conf}(X_1,X_2,\Ko)\right.\\&\left.
\times C_{\conf}(\as)\>\cA^{\PT}_{\conf}\left(\sh,Q_t,\Ko\right)\right\},
\end{split}
\end{equation}
with
\begin{equation}
  \label{eq:sigQm}
  \sigma(Q_m)=\int_{Q_m}^{Q_M}dQ_t 
\int dX_1dX_2\sum_{f\conf}
\left\{\frac{d\hat \sigma^{f}_{\conf}}{dQ_t}
\, \cP^{f}_{\conf}(X_1,X_2,K)\right\}.
\end{equation}
The exact value of the hard scale of $\cP$ in \eqref{eq:sigQm} is not
important, as long as of order $\sqrt{\sh}$; a variation can be
absorbed into a modification of the coefficient $C_{\conf}$.  In order
to simplify the calculation of $C_{\conf}$ for the exact fixed order
expression of $\Sigma$ in \eqref{eq:sigQm} we have fixed this scale at
the same value $K$ in \eqref{eq:rab3} at which the radiation factor
\eqref{eq:sigPT=1} is one.

\subsection{Matching resummed with fixed order results}
In the matching procedure one starts by determining the coefficients
of
\begin{equation}
  \label{eq:Cexp}
  C_{\conf}(\as)=1+\frac{\as}{2\pi}c_1+\cdots 
\end{equation}
from the fixed order exact results. Since only the first loop term
\begin{equation}
\label{eq:sigPT-exact}
\begin{split}
\Sigma^{\exact}=1+\frac{\as}{2\pi}\>\Sigma_1^{\exact}+\ldots\\
\end{split}
\end{equation}
is known we can determine only $c_1$.   
The first term  $\Sigma_1^{\exact}$, which has the DL and SL structure
\begin{equation}
  \label{eq:Sig1}
\Sigma_1^{\exact}(Q_m,\Ko)=-2C_T\,L^2+G_{11}(Q_m)\,L+C_1\>,
\qquad L\equiv\ln\frac{M}{\Ko}\>,
\end{equation}
is obtained\footnote{Since $K$ in the radiator \eqref{eq:Rad-PT} may
  depend on the integration variables, the logarithmic variable for
  the full distribution $\Sigma$ is chosen to be $L$ and not $L_K$. We 
  need then to take into account the mismatch between $L$ and $L_K$.}
by using the numerical program DYRAD~\cite{DYRAD}.  First
of all we check that the DL and SL terms coincide with the result of
our calculation, see \eqref{eq:G11}.
Then we compute $c_1$ which is given by
\begin{equation}
  \label{eq:C1}
  c_1(Q_m,\Ko)=-r_1(Q_m)+
\left\{\Sigma_1^{\exact}(Q_m,\Ko)+2C_T\,L^2\>-\>G_{11}(Q_m)\,L\right\},
\end{equation}
with $r_1$ coming from the mismatch between $L_K$ and $L$, see
\eqref{eq:GC} and \eqref{eq:G11}. The logarithmic terms are completely
subtracted and $c_1$ is finite for $\Ko\to0$.  This is shown in
Fig.~\ref{fig:C1} for some values of $Q_m$.  For simplicity here we
set the hard scale $K$ at $M$ so that $L_K\to L$ and $r_1(Q_m)\to0$.

\EPSFIGURE[ht]{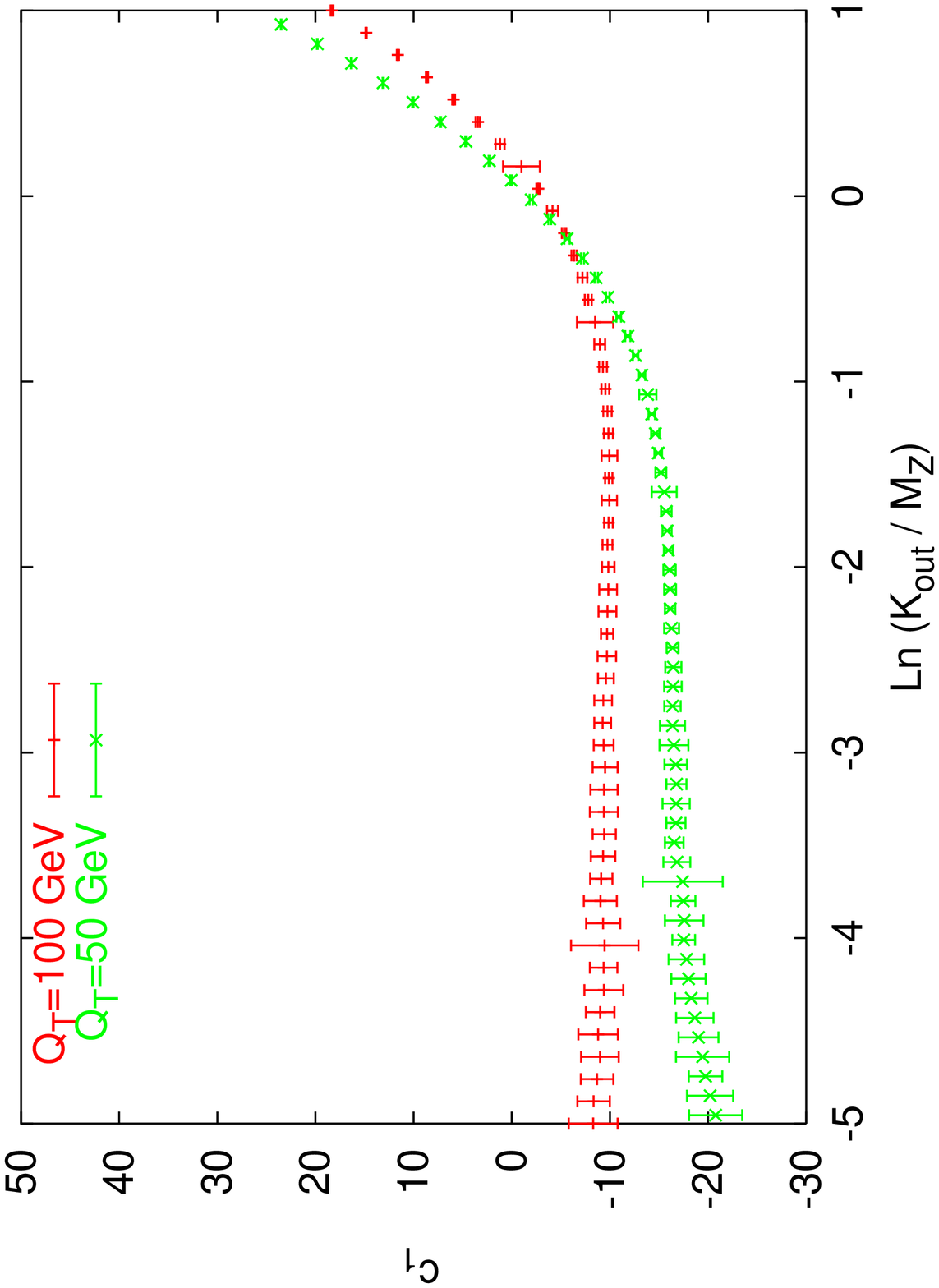,width=0.5\textwidth,angle=270} {The
  first order coefficient function $c_1$ as a function of $\Ko$ for
  the two values $Q_m=50,100\GeV$ and $K=M$.
\label{fig:C1}}

Given $c_1$ in \eqref{eq:C1} we now reconstruct the matched result.
For instance, using the so called ``log-R matching prescription'', to
first order we have
\begin{equation}
\label{eq:ClogR}
\left[\,C_{\conf}(\as)\,\right]^{\mat}_1 \>=\>e^{\frac{\as}{2\pi}c_1}\>.
\end{equation}
It is straightforward to check that the distribution $\Sigma^{\PT}$
obtained by using this coefficient function reproduces the exact
result in \eqref{eq:sigPT-exact} and accounts for all terms of the
form $\as^n L^{2n-2}$ in the resummed expression
\eqref{eq:SigPT-fine}.  In order to obtain also the $\as^n L^{2n-3}$
terms one should perform a second order matching, which requires the
fixed order exact result to order $\as^2$, not yet available.

Actually this expression for the distribution $\Sigma$ is not yet
normalized to one at the maximum value $\Ko^M$ of $\Ko$. 
Indeed we used a normalization point at $\Ko=K$, see \eqref{eq:sigPT=1}.
The standard way to achieve the correct normalization is to substitute 
\begin{equation}
  \label{eq:tKo}
\begin{split}
\frac{K}{\Ko}\>\to\> 
\frac{K}{\tKo}\equiv \frac{K}{\Ko}-\frac{K}{\Ko^M}+1\>,
\qquad L\>\to\>\tilde L\equiv\ln\frac{M}{\tKo}\>,
\end{split}
\end{equation}
in all previous expressions (except in $\Sigma_1^{\exact}(Q_m,\Ko)$ in
\eqref{eq:C1} which is already correctly normalized).  The variable
$\Ko$ goes to the correct kinematical boundary ($\Ko\to\Ko^M$) for
$\tKo\to K$ and tends to $\tKo$ for small values.

\subsection{Including the NP correction}
We now discuss the full distribution including the leading NP
corrections coming from the radiation factor. Recall that power
corrections from $\cP$ are subleading.  The analysis is similar to the
one in \cite{Tmin}.  The NP radiator $\de R_{\conf}$ is proportional
to the Mellin variable $\nu$, see \eqref{eq:Rad-NP}, and so it
produces a {\it shift} of $\Ko$
\begin{equation}
  \label{eq:shift-fine}
   \Ko \to \Ko'\equiv \Ko-\de\Ko\,\qquad
   \de\Ko=\cp\,\Delta_{\conf}(\Ko)\>.
\end{equation}
The final expression, including NP corrections, is then obtained from
the PT result of previous subsections in which we replace $\Ko$ with
$\Ko'$.

Here $\cp$ is expressed in terms of the NP parameter $\al_0(\mu_I)$,
see \eqref{eq:cp}.  To evaluate $\Delta_{\conf}$ we observe that
\eqref{eq:Rad-NP} contains a $\ln |q_{3x}|$ term. The $q_{3x}$
distribution is given by the radiation factor \eqref{eq:hatsigma}
which leads to $|q_{3x}|\sim\Ko$, so that a $\ln|q_{3x}|$ term
produces a $\ln\Ko$ contribution. We find (see Appendix
\ref{App:NPdist})
\begin{equation}
  \label{eq:Delta}
\begin{split}
\Delta_{\conf}(\Ko)&=C^{(\conf)}_1(\eta_0-\eta_3)
+C^{(\conf)}_2(\eta_0+\eta_3)+C^{(\conf)}_3\,\ln\frac{Q^{\NP}}{\Ko}\>,\\
\ln Q^{\NP}&=\ln \zeta Q_t +\gam_E
+\psi\left(1+R'\right)
+\half\psi\left(1+\frac{R'}{2}\right)
-\half\psi\left(\frac{1+R'}{2}\right),
\end{split}
\end{equation}
with $R'$ given in \eqref{eq:cF}.  Notice that, expanding in powers of
$R'$, the $p_3$ contribution is
\begin{equation}
\label{eq:Delta1}
 \ln \frac{Q^{\NP}}{\Ko}=\ln\frac{2\zeta Q_t}{\Ko}+\cO{R'}\>.
\end{equation}
The factor $2$ here is simply due to the fact that $p_3$ acquires a
recoil which is equal to $\Ko/2$ for small $R'$.

The effect of the substitution in \eqref{eq:shift-fine} is a
deformation of the PT distribution. First of all the quantity
\eqref{eq:Delta} depends logarithmically on $\Ko$ (both explicitly and
through the SL function $R'$). This implies that the PT curve is
shifted by an amount which decreases with increasing $\Ko$. Moreover,
$\Delta_{\conf}(\Ko)$ depends also on the rapidity $y$ distributed
according to the incoming parton distributions $\cP_{\conf}(\Ko)$.

Here the situation is different from the case of broadening
\cite{broad} or $T_m$ \cite{Tmin} in $\ee$ annihilation in which one
obtains very singular contributions to the shift of order
$1/\sqrt{\as}$.  The difference is due to the different kinematical
situations.  In the two $\ee$ cases one has to consider contributions
in which some hard partons are forced to stay in the event plane. Then
the PT distribution is given by a Sudakov form factor and then the
$1/\sqrt{\as}$ contribution comes from the integration of the
logarithmic term in the hard parton recoil.  For the present
observable instead, $p_3$ is not kinematically forced into the event
plane and then its PT radiation factor reproduces a logarithmic
contribution.

\subsection{Numerical analysis \label{sec:Numerical}}
We report here some numerical results. We consider $p\bar p$ collisions
at Tevatron energy ($\sqrt{s}=1.8$TeV) for some typical values of $Q_m$.  
Data on the $\Ko$ distribution are not yet available.
The results depend on the two parameters $\al_{\MSbar}(M_Z)$ and
$\al_0(\mu_I)$ (with $\mu_I=2\GeV$) which values we fix in the range
determined by the $2$-jet shape analysis~\cite{GZ}. 

\EPSFIGURE[ht]{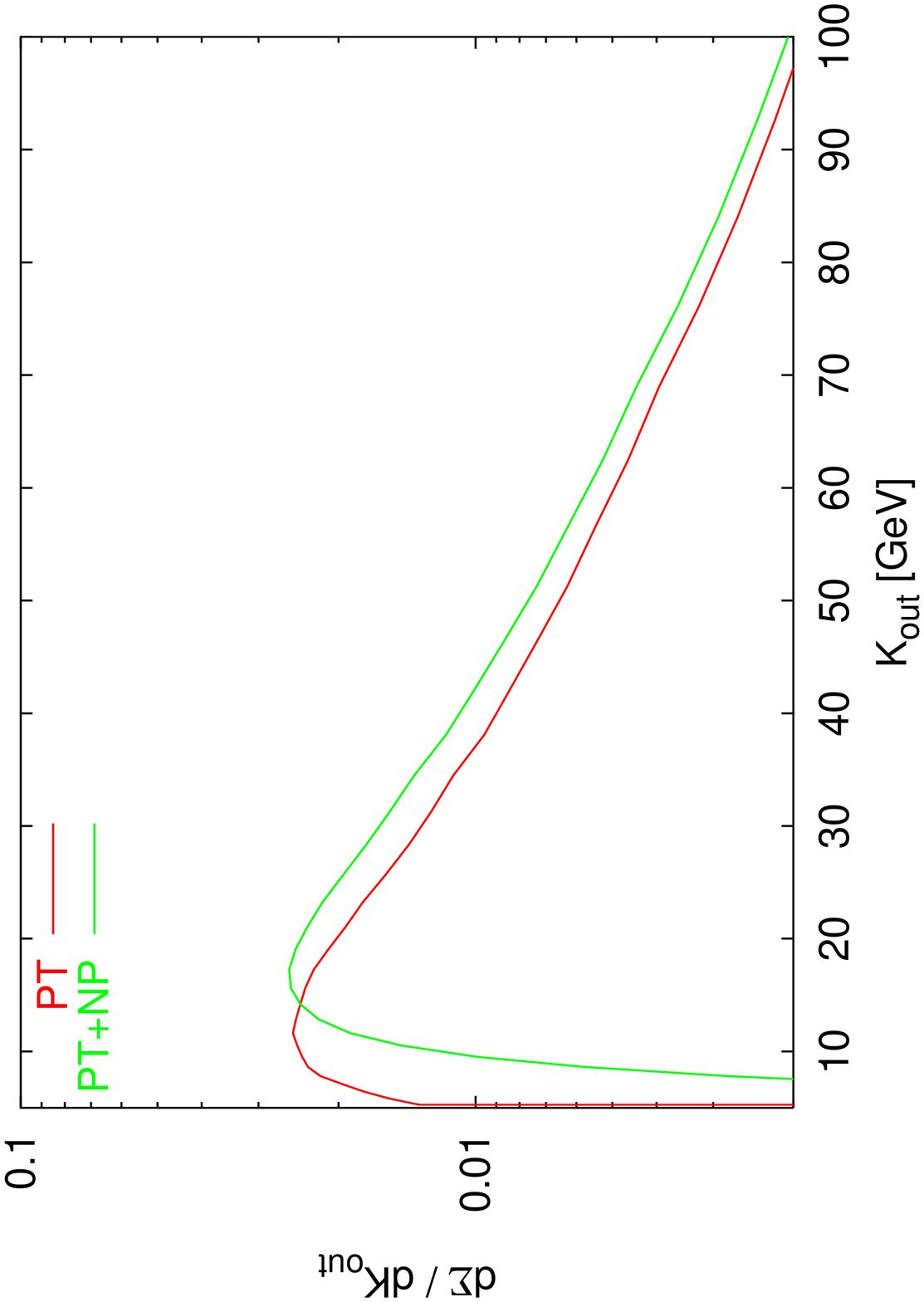,width=0.5\textwidth,angle=270} {The
  differential distribution in \eqref{eq:dSig-num} as function of
  $\Ko$. We plot both the PT resummed and matched result and the full
  distribution including NP correction for $Q_m=50\GeV$.  Here we have
  taken $\al_{\MSbar}(M_Z)=0.120$ and the NP parameter
  $\al_0(2\GeV)=0.52$. The contribution due to the beam remnant
  interaction is not included, it is given by a rigid shift in $\Ko$
  by $\Ko^{\remnant}$ of the order of few GeV, see \eqref{eq:Krem}.
\label{fig:Dist50}}

In figs.~\ref{fig:Dist50}, \ref{fig:Dist100} we plot the differential 
distributions
\begin{equation}
  \label{eq:dSig-num}
 \frac{d\Sigma(Q_m,\Ko)}{d\Ko}\>, 
\end{equation}
for two values of $Q_m$.  
The PT curve represents \eqref{eq:SigPT-fine} with the coefficient
function given by \eqref{eq:ClogR}. We have performed the substitution
\eqref{eq:tKo} in order to take into account the correct normalization
at the kinematical boundary. The rapidity cut is set at $\eta_0=2$.
For simplicity we have taken the hard scale $K$ at $M$. We used the
incoming parton distribution of \cite{SF}. 

The PT$+$NP curve is given by the above PT expression in which we make
the substitution $\Ko$ into $\Ko'$ according to \eqref{eq:shift-fine}.
The leading NP correction is determined by the single parameter
$\al_0(\mu_I)$.

As discussed before, the effect of the NP substitution
\eqref{eq:shift-fine} is a deformation of the PT distribution. In
particular the PT peak is shifted by about $10$ GeV.  This effect, due
to the QCD running coupling in the infrared region, has to be
contrasted with the contribution from the beam remnant interaction
which corresponds to a ``rigid shift'' of the hard QCD result by an
amount $\Ko^{\remnant}$ of order of few GeV, see
\cite{MW,MCworkshops,MW88}, proportional to $\eta_0$ but independent
of the hard scales (see \eqref{eq:Krem}).

\EPSFIGURE[ht]{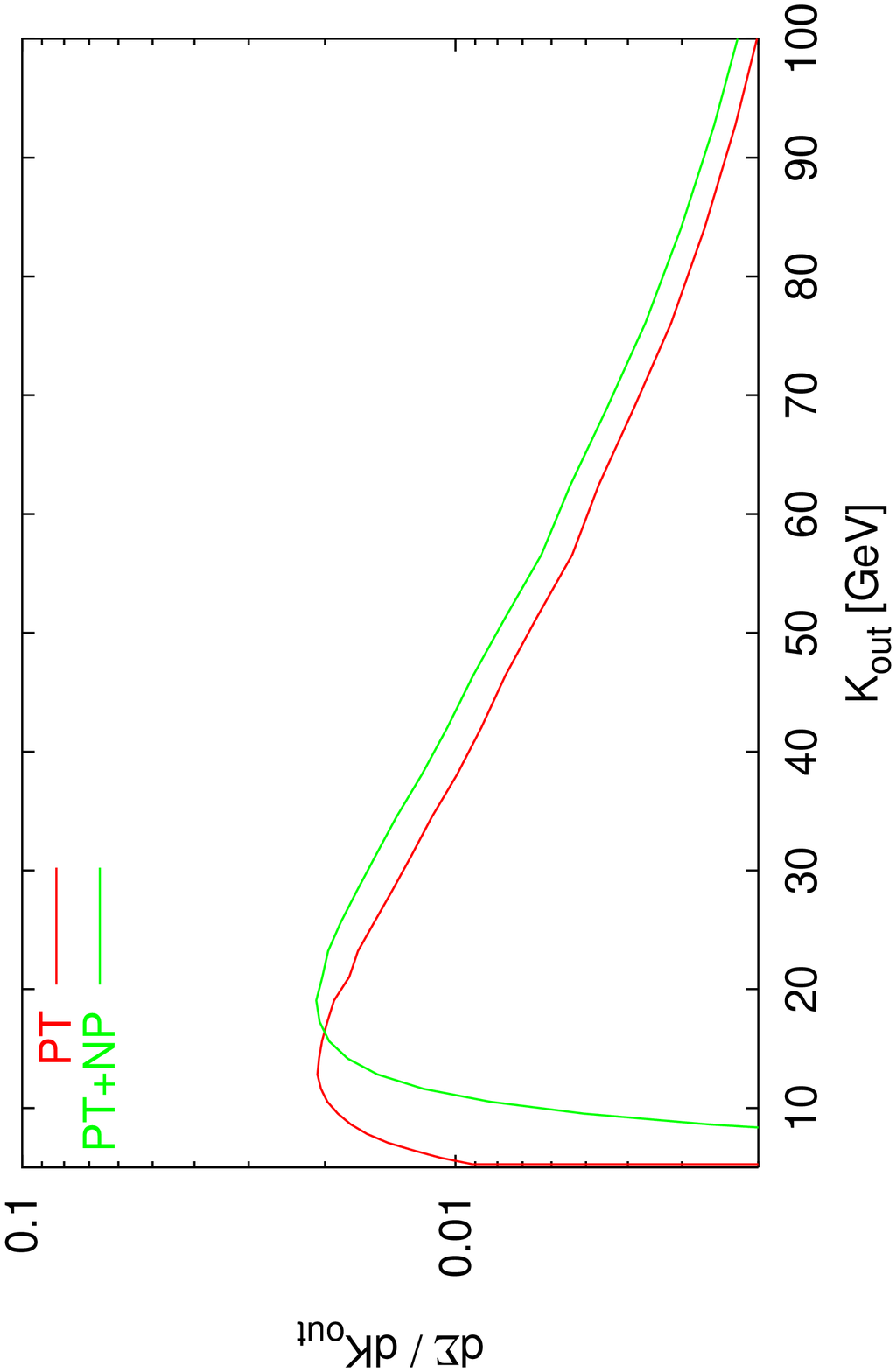,width=0.5\textwidth,angle=270} {Same as 
fig.~\ref{fig:Dist50} for $Q_m=100\GeV$.  
\label{fig:Dist100}}

\section{Discussion and conclusion \label{sec:Discussion}}
The aim of the present study is the understanding of the structure of
radiation in hard hh-collisions within the standard QCD treatment.  We
have introduced the jet-shape observable $\Ko$ which is the extension
to hh-collisions of $T_m$ in $\ee$ annihilation.  
The hard QCD analysis includes next-to-leading order PT resummation,
matching with exact first order result, and leading NP power corrections
(arising from the fact that the running coupling argument runs into
the infrared region).  
To avoid measurements inside the beam direction we have included a
rapidity cut $\eta_0$ (see \eqref{eq:rapidity}). For values of $\Ko$
much smaller than $\Ko^c$ (of the order of $10$ GeV for $\eta_0=2$)
there are powers of $\ln\Ko/\Ko^c$. In our calculation we consider the
region $\Ko>\Ko^c$ so that we do not need to resum them.  As a result
the PT contribution does not depend on $\eta_0$. The dependence on
$\eta_0$ enters only in the NP correction which corresponds to the
substitution \eqref{eq:shift-fine} in the argument of the PT
distribution. 

We discuss now some of the features of our hard QCD result.

In the present calculation we have various hard scales.  It is then
important to identify the specific hard scales in the various
factorized pieces of the result \eqref{eq:SigPT-fine}.  We find that
$\Ko$ is the hard scale for the incoming partons distribution
$\cP_{\conf}$ while it is the lower bound for frequencies contributing
to the radiation factor $\cA_{\conf}$. This result is based on the
different r\^ole of real/virtual cancellations in the collinear
singular quantity $\cP_{\conf}$ and in the CIS radiation factor
$\cA_{\conf}$.  Technically, in the present treatment, the
factorization ($\cP_{\conf}\cdot\cA_{\conf}$) results from
\eqref{eq:split} while real/virtual cancellation from \eqref{eq:rho}.
The hard scales for $\cA_{\conf}$ are given by $Q_a^{(\conf)}\sim
\sqrt{\sh}$, the scales of the elementary hard process. They are
identified at SL level and depend on the geometry of the event ($\sh$-
and $Q_t$-dependence) and on the configuration $\conf=\qq, qg,gq$ for
the two incoming partons. Therefore, the shape in $\Ko$ of the
distribution (at fixed $s$ and $Q_m$) depends on the weight of the
various configurations. By changing $s$ and $Q_m$ one may be able to
study the three configurations separately.

Leading power corrections come from the radiation factor (power
corrections in the incoming parton distributions are subleading
\cite{DMW} and were not considered).  They enter as a shift in the PT
distribution given by \eqref{eq:shift-fine}. This is a feature common
to all observables linear in the transverse momentum.  The strength of
the power correction is given by the NP parameter $\cp$ (expressed in
terms of $\al_0(\mu_I)$, see \eqref{eq:cp}), the same as introduced
for the $\ee$ jet-shape observables. The structure of the shift is
characteristic of the fact that the observable $\Ko$ is uniform in
rapidity, see the discussions after \eqref{eq:eta3} and
\eqref{eq:Delta1}.
The NP shift is larger then the corresponding NP shift for 2-jet
observables since it takes contributions from three hard partons, one
of which is a gluon. This was also the case for the near to planar $3$-jet
observables in $\ee$ \cite{Tmin,Dpar}.  We then expect that higher
order NP effects may come into play near the peak of the distribution.
This calls for a deeper analysis that would address higher power
corrections, for example, along the lines of Korchemsky-Sterman
approach which was recently developed for some 2-jet observables in
\cite{KS}.  The comparison with experimental data (not yet available)
would shed light on this important point.

Numerical programs for exact results on the matrix elements for the
process \eqref{eq:process} are available \cite{DYRAD} to order $\as^2$.
Thus we have been able to compute only the first term of the
coefficient function $C_{\conf}(\as)$.

In $\ee$ annihilation the hard QCD analysis described above has been
shown to be sufficient to make quantitative predictions and study the
universality of NP effects. In hh-collisions however we also need to
take into account contributions coming from the soft underlying event
due to beam remnant interaction. 
With the hypothesis that such contributions are factorized and
independent of the hard scales (see \cite{MW,MCworkshops,MW88} and
Appendix \ref{App:beam}), they give rise to a ``rigid shift'' in $\Ko$
by the amount $\Ko^{\remnant}$ (see \eqref{eq:Krem}).  This quantity,
of the order of few GeV, is proportional to $\eta_0$ and the minimum bias
parameters $\cN$ and $\VEV{k_t}^{\remnant}$. The last two parameters
should be the same in all hard hh-processes and could be determined
and checked in the study of various observables. 

This completes the analysis of our jet-shape observable in hard
hh-collisions. The results depend on the two parameters $\as(M_Z)$ and
$\al_0(\mu_I=2\GeV)$. The important question is then whether the
present QCD standard treatment is sufficient to reproduce the data in
hard hh-collisions.  In particular, our results should provide solid
ground to study whether there is any need for non-hard QCD
contributions.

\section*{Acknowledgements}
We are grateful to Yuri Dokshitzer, Nigel Glover, Michelangelo Mangano,
Gavin Salam and Bryan Webber for helpful discussions and suggestions.

\newpage 

\appendix
\section{Elementary cross sections \label{App:Born}}
We present the three elementary cross sections relating to the
configurations $\qq$, $qg$ and $gq$ in terms of the kinematical
variables $\sh$ and $\theta$ introduced in equation \eqref{eq:Ekin}.

The parton-level cross section $\sig_\delta^f$ for quark flavour 
$f$ and configuration $\delta$ is
\begin{equation}
\sig_\delta^f = \frac{1}{2\sh}\int d\mbox{Lips}[P_1 P_2\to QP_3]
\cdot |M_\delta^f|^2 \>,
\end{equation}
where the Lorentz-invariant integration measure, after eliminating
a trivial azimuthal angle, becomes simply
\begin{equation}
\int d\mbox{Lips}[P_1 P_2\to QP_3] = 
\frac{\sh-M^2}{16\pi\sh}\int d\cos\theta \>.
\end{equation}

For the matrix elements $M_\delta^f$ only the simple $2\to2$
tree-level diagrams are required. After averaging over colours
and spins we obtain
\begin{equation}
\begin{split}
&|M_{\qq}^f|^2 = \frac{32\pi^2\as\al_f C_F}{N_c}
\frac{(P_1Q)^2+(P_2Q)^2}{(P_1P_3)(P_2P_3)} \\
&|M_{qg}^f|^2 = \frac{32\pi^2\as\al_f T_R}{N_c}
\frac{(P_1Q)^2+(P_3Q)^2}{(P_1P_2)(P_2P_3)} \\
&|M_{gq}^f|^2 = \frac{32\pi^2\as\al_f T_R}{N_c}
\frac{(P_2Q)^2+(P_3Q)^2}{(P_1P_2)(P_1P_3)} \>,
\end{split}
\end{equation}
where $\al_f$ is the electroweak coupling of the quark to the $Z_0$
\begin{equation}
4\pi\al_f = \left(\frac{g_W}{2\cos\theta_W}\right)^2
\left(V_f^2+A_f^2\right)\>.
\end{equation}

Using \eqref{eq:Ekin} to express these matrix elements in terms 
of $\sh$ and $\theta$ quickly yields the differential cross
sections
\begin{equation}
  \label{eq:hard}
\begin{split}
\frac{d\hat \sigma^{f}_{\qq}(\sh,\theta)}{d\cos\theta}
&=\>\frac{2\pi\as \al_f C_F}{N_c}\cdot
\frac{2(E^2+p^2\cos^2\theta)}{p\,\sh^{3/2}\sin^2\theta}\>,\\
\frac{d\hat \sigma^{f}_{qg}(\sh,\theta)}{d\cos\theta}
&=\>\frac{2\pi\as \al_f T_R}{N_c}\cdot
\frac{(E-p\cos\theta)^2+4p^2}{\sh^2(1-\cos\theta)}\,\\
\frac{d\hat \sigma^{f}_{gq}(\sh,\theta)}{d\cos\theta}
&=\>\frac{2\pi\as \al_f T_R}{N_c}\cdot
\frac{(E+p\cos\theta)^2+4p^2}{\sh^2(1+\cos\theta)}\,
\end{split}
\end{equation}

We however are interested in $d\hat{\sigma}_\conf^f/dQ_t$, which is given by
\begin{equation}
\label{eq:dQt}
\frac{d\hat{\sigma}_\conf^f(\sh,Q_t)}{dQ_t} = \int_{-1}^1 d\cos\theta\,
\delta(Q_t-p\sin\theta)\,
\frac{d\hat{\sigma}_\conf^f(\sh,\theta)}{d\cos\theta}\>.
\end{equation}
On integrating over $\theta$ there are contributions
from $\cos\theta = \pm\sqrt{1-Q_t^2/p^2}$. Thus
\begin{equation}
\label{eq:dQt1}
\frac{d\hat{\sigma}_\conf^f(\sh,Q_t)}{dQ_t} = 
\left.\frac{\tan\theta}{p}\frac{d\hat{\sigma}_\conf^f}{d\cos\theta}
\right|_{\cos\theta>0} -
\left.\frac{\tan\theta}{p}\frac{d\hat{\sigma}_\conf^f}{d\cos\theta}
\right|_{\cos\theta<0} \>.
\end{equation}
Now we make use of the fact that, for proton-proton and proton-antiproton
collisions, these two terms contribute equally to \eqref{eq:dsigmaB}.
(This would not of course be true for collisions of two arbitrary hadrons.)
Therefore we may take $\cos\theta>0$ and write
\begin{equation}
\label{eq:dQt2}
\frac{d\hat{\sigma}_\conf^f(\sh,Q_t)}{dQ_t} = 2\frac{\tan\theta}{p}
\frac{d\hat{\sigma}_\conf^f(\sh,\theta)}{d\cos\theta}\>,\quad\quad
\sin\theta = Q_t/p \>,\quad
\cos\theta = \sqrt{1-Q_t^2/p^2}\>.
\end{equation}

The relation $\sin\theta=Q_t/p$ gives a lower bound on the total
centre of mass energy $\sqrt{\sh}$:
\begin{equation}
  \label{eq:sh0}
\sqrt{\sh}>\sqrt{\sh_0}=Q_t+\sqrt{M^2+Q_t^2}\>.
\end{equation}
Finally, a consideration of the kinematics shows that, for fixed partonic
energy $\sqrt{\sh}$ the observable is bounded by
\begin{equation}
  \label{eq:Koumax}
  \Ko^2<\sh+M^2-2\sqrt{\sh(Q_t^2+M^2)}\>,
\end{equation}
and since $Q_t$ can be as low as $Q_m$ and $\sh$ can (with small probability)
be as large as $s$, we have the absolute upper bound
\begin{equation}
\label{eq:K_M}
K_M^2=s+M^2-2\sqrt{s(Q_m^2+M^2)}\>.
\end{equation}

\section{Incoming parton evolution at the hard scale \label{App:Gamm}}
We denote by $s_{ab}$ the contribution to $\Gamma_{\conf}$ from a
given $ab$-dipole distribution $w_{ab}$, see \eqref{eq:W}, and
consider first the $12$-dipole contribution
\begin{equation}
  \label{eq:s12}
s_{12} = \sum_{a=1}^{2} 
\int_{\cC_{a}} \frac{d^3k}{\pi\om}w_{12}(k)(1-z^{N_a-1})\,u(k_x)
=\sum_{a=1}^2 S_1(N_a)\int\frac{d^2k_t}{\pi k_t^2}\frac{\as(k_t)}{\pi}\,
e^{-\nu(|k_x|-i\be k_x)}\>,
\end{equation}
with 
\begin{equation}
  \label{eq:S1}
S_1(N)=\int_0^1 dz \;\frac{1-z^{N-1}}{1-z}=\sum_{n=1}^{N-1}\frac{1}{n}\>,
\end{equation}
proportional to the soft piece of the anomalous dimension. Here both
collinear regions $\cC_1,\cC_2$ contribute.  

The contribution from the other two radiators $s_{13},s_{23}$ is
similar. They take a contribution from a single collinear region
(the $a3$-dipole from the region $\cC_a$ collinear to $p_a$), so that
\begin{equation}
  \label{eq:sa3}
s_{a3} =S_1(N_a)\int\frac{d^2k_t}{\pi k_t^2}\frac{\as(k_t)}{\pi}\,
e^{-\nu(|k_x|-i\be k_x)}\>,\qquad a=1,2\>.
\end{equation}

Since the $s_{ab}$ are singular for $k_x\to 0$, one needs to introduce
a cutoff $\mu$ on the $k_t$ integral. We also use the fact that, to
our required accuracy, the $k_x$-source can be replaced by an
effective cutoff (see \eqref{eq:rho}).
We are therefore required to integrate over a region in $k_x,k_y$-space
given by $\Theta(\rho^{-1}-|k_x|)\Theta(k_t-\mu)$, where we choose $\mu$
to be less than $\rho^{-1}$. So, to SL accuracy, we write (for instance
for the $a3$-dipole)  
\begin{equation}
  \label{eq:sint1}
s_{a3}=S_1(N_a)\int_{\mu}^{1/\rho}\frac{dk_t}{k_t}\frac{2\as(k_t)}{\pi}\>,
%+ \Omega_{a3} \>,
\end{equation}
where the remainder is beyond our required SL accuracy.

As anticipated, since here one has $|k_x|<\rho^{-1}\ll Q_t$, the 
precise definition of the collinear regions $\cC_1,\cC_2$ is not
important.

\section{The PT radiator \label{App:RadPT}}
The PT radiator is given, to SL accuracy, in terms of $ab$-dipole
radiators
\begin{equation}
  \label{eq:rab}
r_{ab}(\nu,\be)=\int\frac{d^3k}{\pi\om}w_{ab}(k)\,
\left[1-e^{-\nu(|k_x| -i\be k_x)}\right],
\qquad w_{ab}(k)=\frac{\as(k^2_{ab,t})}{\pi k^2_{ab,t}}\>,  
\end{equation}
where $k_{ab,\,t}$ is the invariant transverse momentum of $k$ with
respect to the $P_a,P_b$ hard partons in \eqref{eq:Ekin}.  For the
configuration $\conf=\qq$, for instance, we have
\begin{equation}
  \label{eq:R12}
R^{\PT}_{\qq}(\nu,\be)=\frac{N_c}{2}\left(r_{13}(\nu,\be)+r_{23}(\nu,\be)
-\frac{1}{N_c^2}\,r_{12}(\nu,\be)\right).
\end{equation}
To evaluate the $ab$-dipole radiator $r_{ab}(\nu,\be)$ we work in the
centre of mass system of the $ab$-dipole. We neglect at this stage the
rapidity cut \eqref{eq:rapidity}: we will show at the end of this appendix
that the difference is beyond our accuracy. Denoting by $P^*_a,P^*_b$
and $k^*$ the momenta in this system, we introduce the Sudakov
decomposition
\begin{equation}
  \label{eq:Sud}
     P_a^*=\frac{Q_{ab}}{2}(1,0,0,1)\>,\quad 
     P_b^*=\frac{Q_{ab}}{2}(1,0,0,-1)\>,
\qquad k^*=\al P_a^*+\be P_b^*+\ka\>, 
\end{equation}
where $Q_{ab}^2=2(P_aP_b)$ so that \eqref{eq:Ekin} gives
\begin{equation}
  \label{eq:Q_ab}
Q^2_{12}=\sh\>,\quad   
Q^2_{13}=p\sqrt{\sh}\,(1+\cos\theta)\>,\quad
Q^2_{23}=p\sqrt{\sh}\,(1-\cos\theta)\>,
\end{equation}
and $\theta$ is given in equation \eqref{eq:dQt2}.
Here the two-dimensional vector $\vka$ is the transverse momentum
orthogonal to the $ab$-dipole momenta ($\ka^2=k^2_{ab,t}$). We have
then
\begin{equation}
  \label{eq:ab-cm}
w_{ab}(k)=\frac{\as(\ka^2)}{\pi \ka^2}\>, \qquad
\frac{d^3k}{\pi\omega}=\frac{d^2\ka}{\pi}\frac{d\al}{\al}\>, \quad 
\al>\frac{\ka^2}{Q^2_{ab}}\>.
\end{equation}
Since, neglecting the recoil $q_{3}$, the outgoing momentum $P_3$ is
in the $yz$-plane, the Lorentz transformation is in the $yz$-plane and
our observable $k_x$ remains unchanged. The $ab$-radiator has then the
form
\begin{equation}
  \label{eq:rab1}
  r_{ab}(\nu,\be)=2\int_0^{Q^2_{ab}}\frac{d^2\ka}{\pi\ka^2}
\frac{\as(\ka)}{\pi} \int_{\ka/Q_{ab}}^1\frac{d\al}{\al}
\left[1-e^{-\nu(|k_x|-i\be k_x)}\right],
\quad \ka_x=k_x\>,
\end{equation}
where the factor $2$ comes because we have integrated only over the
``right hemisphere'' $\al>\ka/Q_{ab}$.  Integrating over $\al$ and
$\ka_y$ we have
\begin{equation}
 \label{eq:rab2}
r_{ab}(\nu,\be)=
2\int_{-Q_{ab}}^{Q_{ab}}\frac{dk_x}{|k_x|}
\left[1-e^{-\nu(|k_x|-i\be k_x)}\right]
\cdot \frac{\as(2|k_x|)}{\pi}\ln\frac{Q_{ab}}{2|k_x|}\>.
\end{equation}
To show this we introduced $\ka_y=t\cdot |k_x|$ and used
\begin{equation}
\int_{-\infty}^{\infty}\frac{dt}{\pi(1+t^2)}\as\left(|k_x|\sqrt{1+t^2}\right)
\ln \frac{Q_{ab}}{|k_x|\sqrt{1+t^2}}\simeq
\as(2|k_x|)\ln\frac{Q_{ab}}{2|k_x|}\>.
\end{equation}
We extended the $t$-integration to infinity since it is
convergent, then we integrated over $t$ by expanding $\as$ to
second order. Corrections are beyond SL accuracy. Finally, using
\eqref{eq:rho}, we obtain
\begin{equation}
  \label{eq:rab4}
r_{ab}(\nu,\be)=4\int_{\rho^{-1}}^{Q_{ab}}
\frac{dk_x}{k_x}\frac{\as(2k_x)}{\pi}\>
\ln\frac{Q_{ab}}{2k_x}\>.
\end{equation}
Assembling the various dipole contributions and including hard
collinear splittings then yields, to SL accuracy, \eqref{eq:Rad-PT}
and \eqref{eq:rab3}.

We now show that the rapidity cut \eqref{eq:rapidity} is negligible
for the PT radiator. The difference between the radiator with the cut
imposed and that without is given by the dipole contributions
\begin{equation}
\label{eq:Raddif}
\Delta r_{ab}(\nu,\be)=-\int_{|\eta_k|>\eta_0}\frac{d^3k}{\pi\om}
\>w_{ab}(k)\>e^{i\nu\be k_x}\left(1-e^{-\nu|k_x|}\right)\>.
\end{equation}
In order to implement the rapidity cut, we express $\eta_k$ the soft
gluon rapidity in the frame \eqref{eq:hhq} in the invariant form:
\begin{equation}
\label{eq:eta_k}
\eta_k = \frac{1}{2}\ln\frac{(h_2k)}{(h_1k)} 
= y + \frac{1}{2}\ln\frac{(P_2k)}{(P_1k)} \>,
\end{equation}
where $h_1,h_2$ are the incoming hadron momenta in \eqref{eq:hhq}
and $P_1,P_2,y$ are the hard incoming parton momenta and rapidity in
\eqref{eq:Ekin}.

For the $12$ dipole we have $\eta_k=y+\ln(\al Q_{12}/\ka)$, and thus
we obtain from the ``right'' hemisphere
\begin{equation}
\label{eq:Raddif1}
\Delta r_{12}(\nu,\be)=-\int_0^{Q_{12}'}\frac{d^2\ka}{\pi\ka^2}
\frac{\as(\ka)}{\pi}\int_{\ka/Q_{12}'}^1\frac{d\al}{\al}
\>e^{i\nu\be k_x}\left(1-e^{-\nu|k_x|}\right)\>,\quad
Q_{12}'=Q_{12}e^{y-\eta_0}\>.
\end{equation}
Here the scale for the correction is $Q_{12}'\sim\Ko^c\le\cO{\nu^{-1}}$ 
and so
the contribution is of order $\as(\Ko)$ without a logarithmic enhancement.
The same is found for the ``left'' hemisphere contribution.

For the $13$ dipole we obtain a similar result for the cut around the
$P_1$ direction, (using \eqref{eq:13rap} to evaluate $\eta_k$),
while the cut around the $P_2$ direction gives a tiny
correction proportional to the size of the hole: the converse holds for
the $23$ dipole.

\section{NP corrections to the radiator \label{App:RadNP}}
We consider the NP correction $\de r_{ab}$ to the $ab$-dipole
radiator.  In this case, as we shall see, we need to retain both the
recoil $q_{3x}$ and the rapidity cut $\eta_0$ (recall that in the PT
component they both gave subleading effects and were neglected). We
write the integral in the $ab$-dipole centre of mass variables
$\al,\be$ and $\vka$ introduced in \eqref{eq:Sud} and, to obtain the
NP correction $\de r_{ab}$, we perform the following standard
operations:
\begin{itemize}
\item the running coupling, reconstructed by two loop emission, is
  represented by the dispersive form \cite{DMW}.  Then, the
  $ab$-dipole radiation $w_{ab}(k)$ is written in the $ab$-centre of
  mass system (see \eqref{eq:ab-cm}) in the form
  \begin{equation}
    w_{ab}(k)=\frac{\as(\ka)}{\pi \ka^2}=\int_0^{\infty} 
    \frac{dm^2\,\aef(m)}{\pi(\ka^2+m^2)^2}\>;
  \end{equation}
\item to take into account the emission of soft partons at two loop
  order \cite{Milan}, we need to extend the source $u(k_x)$ to include
  the mass $m$ of the soft system.  We assume
  $k_x=\ka\cos\phi\to\sqrt{\ka^2+m^2}\cos\phi$, with $\phi$ the
  azimuthal angle of $\vka$. Similarly we introduce the mass in the
  kinematical relations such as $\al\be=(\ka^2+m^2)/Q_{ab}^2$ for the
  $ab$-dipole variables;
\item we take the NP part $\de\aef(m)$ of the effective coupling.
  Since it has support only for small $m$, we take the leading part
  of the integrand for small $\ka$, and $m$.  In particular we
  linearize the source $U(k)$
  \begin{equation}
\label{eq:linearU}
    \left[1-U(k)\right]\>\to\> \nu\sqrt{\ka^2+m^2}\>|\cos\phi|\>
\Theta(\eta_0-|\eta_k|)\>.
\end{equation}
Recall that $\eta_k$ is the rapidity of $k$ in the laboratory system
\eqref{eq:hhq}. Here we have neglected terms proportional to $\be$ since
they vanish, by symmetry, upon the $\be$ integration;
\item the recoil component $q_{3x}$ of the outgoing parton does
  provide an effective cut in the soft gluon rapidity along the
  outgoing parton \cite{Tmin,broad}. This is due to a real-virtual
  cancellation which takes place when the angle of the outgoing parton
  $p_3$ with the event plane exceeds the corresponding angle of the
  soft gluon. The detailed analysis of real and virtual pieces entails
  that the contribution from the observable
  $\sqrt{\ka^2+m^2}\>|\cos\phi|$ in the linear expansion of the source
  (see \eqref{eq:linearU}) has to be replaced by
\begin{equation}
  \label{eq:coherence}
\sqrt{\ka^2+m^2}\>|\cos\phi|\to
\left|\sqrt{\ka^2+m^2}\,\cos\phi+\al q_{3x}\right|-\al|q_{3x}|\>,
\end{equation}
with $\al$ the Sudakov variable in the $a3$-dipole centre of mass;
\item to take fully into account effects of non-inclusiveness of jet
  observables at two-loop order, we multiply the radiator by the Milan
  factor \cite{Milan,Milan2}
  \begin{equation}
    \label{eq:cM}
  \cM = \frac{3}{64}\frac{(128\pi+128\pi\ln 2-35\pi^2)C_A-5\pi^2
    n_f}{11C_A-2 n_f}\>,
\end{equation}
using $n_f=3$;
\item the NP correction is finally expressed in terms of the parameter
  \begin{equation}
    \label{eq:cp'}
\cp=2\cM\, c_{\Ko}\int dm\frac{\de\aef(m)}{\pi}\>, 
\quad c_{\Ko}=\frac{2}{\pi}\>.
\end{equation}
After merging PT and NP contributions to the observable in a
renormalon free manner, one has that the distribution is independent
of $\mu_I$ and one obtains
\begin{equation}
  \label{eq:cp}
 \cp \!\equiv  c_{\Ko}\,\cM \frac{4}{\pi^2}\mu_I
\left\{ \alpha_0(\mu_I)- \bar{\alpha}_s
  -\beta_0\frac{\bar{\alpha}_s^2}{2\pi}\left(\ln\frac{Q}{\mu_I} 
+\frac{K}{\be_0}+1\right) \right\},  
\end{equation}
where 
\begin{equation}
\bar{\alpha}_s\equiv \al_{\MSbar}(Q)\>,\quad
   K\equiv
  C_A\left(\frac{67}{18}-\frac{\pi^2}{6}\right)-\frac{5}{9}n_f \>,
  \quad \beta_0=\frac{11N_c}{3}-\frac{2n_f}{3}\>.
\end{equation}
The $K$ factor accounts for the mismatch between the $\MSbar$ and the
physical scheme \cite{CMW} and $\al_0(\mu_I)$ is given in
\eqref{eq:al0}. Here $Q$ is the renormalisation scale used in the
next-to-leading order PT calculation.

The numerical coefficient $c_{\Ko}$ depends on our observable $\Ko$.
For instance, the shift for the $\tau=1\!-\!T$ distribution is
\begin{equation}
  \label{eq:thrust}
  \frac{d\sigma}{d\tau}(\tau)=
\frac{d\sigma^{\PT}}{d\tau}(\tau\!-\!\Delta_{\tau})\>,
\qquad   \Delta_{\tau}=C_F\,\frac{c_{\tau}\,\cp}{c_{\Ko}}\>,
\quad c_{\tau}=2\>,
\end{equation}
where $C_F$ enters due to the fact the two-jet system is made of
a quark-antiquark pair.

\end{itemize}
We recall that these prescriptions correspond to taking into account NP
corrections at two-loop order in the reconstruction of the
(dispersive) running coupling and in the non-inclusive nature of
the observable. 
We implement the rapidity cut by expressing $\eta_k$ the soft
gluon rapidity in the invariant form \eqref{eq:eta_k}.

\subsection{Dipole $12$}
This procedure gives, for the $12$-dipole contribution,
\begin{equation}
\label{eq:der12}
\begin{split}  
\de r_{12}
&= \frac{\nu\cM}{\pi}\! \int \! dm^2\de\aef(m)\frac{-d}{dm^2}\!
\int \frac{d\ka^2}{\ka^2+m^2}\int_{-\pi}^{\pi }\frac{d\phi}{2\pi} 
\sqrt{\ka^2+m^2}|\cos\phi|\cdot 2\eta_0\\
&= \nu\,\cp \>2\eta_0\>.
\end{split}
\end{equation}
We used the $12$-dipole centre of mass variables $\al,\be$ and $\vka$
introduced in \eqref{eq:Sud}. This result is found by using
\eqref{eq:eta_k} and
\begin{equation}
\frac{P_2k}{P_1k}=\frac{\al}{\be}=\frac{(\al Q_{12})^2}{\ka^2+m^2}\>,
\end{equation}
so that the $\al$ integration yields
\begin{equation}
  \label{eq:2eta0}
\int_0^1\frac{d\al}{\al}\Theta\left(\eta_0-
\left|\ln\frac{\al Q_{12}e^{y}}{\sqrt{\ka^2+m^2}}\right|\right)\to 2\eta_0\>, 
\qquad \ka,m\to0\>.
\end{equation}
The observable is uniform in rapidity and its integration gives
$2\eta_0$. Corrections coming from the presence of the recoil $q_{3x}$
can be neglected in this case.

\subsection{Dipoles $13$ and $23$}
We consider now the NP corrections $\de r_{a3}$ to the $a3$-dipole
radiators and again we use the $a3$-dipole centre of mass variables
$\al,\be$ and $\vka$ introduced in \eqref{eq:Sud}. The situation is
different from the previous $12$-case in two respects.

First of all we have that the $\eta_0$ cut in the soft gluon rapidity
$\eta_k$ does not affect the region along the hard outgoing parton
$p_3$. There one has to take into account that, as in the case of
broadening \cite{broad} or thrust minor \cite{Tmin}, the recoil component
$q_{3x}$ of the outgoing parton provides an effective cut in the soft
gluon rapidity. If the three momenta in this system are given by the
Sudakov decomposition
\begin{equation}
  \label{eq:Suda3}
     P_3^*=\frac{Q_{a3}}{2}(1,0,0,1)\>,\quad 
     P_a^*=\frac{Q_{a3}}{2}(1,0,0,-1)\>,
\qquad k^*=\al P_3^*+\be P_a^*+\ka\>,
\end{equation}
we are required to use the expression given in \eqref{eq:coherence} as
our linearized source.

The second complication for the $a3$ case is that the expression of
$|\eta_{k}|$ in terms of the variables in \eqref{eq:Suda3} is rather
complex, due to the fact that the Lorentz transformation to go from
\eqref{eq:Ekin} to \eqref{eq:Suda3} involves both a $yz$-rotation and
a boost along the $y$-axis.  For our analysis, we need the expression
for $\eta_k$ only for the soft gluon close to the incoming parton 
direction.

We consider first the case of $13$-dipole. For ${k}$ nearly parallel
to ${P}_1$ we have $\be\gg\al$ and this gives
\begin{equation}
\label{eq:13rap}
\frac{P_2k}{P_1k}\simeq \left(\frac{Q_{12}} {Q_{13}}\right)^2 \cdot
\frac{\ka^2+m^2}{(\al Q_{13})^2}\>,
\end{equation}
The NP correction to the $13$-dipole radiator is then given by
\begin{equation}
\begin{split}
\de r_{13}&=\frac{\nu\cM}{\pi}\!\int\!dm^2\de\aef(m)\frac{-d}{dm^2}\!
\int \frac{d\ka^2}{\ka^2+m^2}\cdot I_{13}\>,\\
I_{13}&=\int_{-\pi}^{\pi }\frac{d\phi}{2\pi} 
\int_{0}^{1}\frac{d\al}{\al}
\left(\left|\sqrt{\ka^2\!+\!m^2}\cos\phi\!+\!\al q_{3x}\right|
\!-\!\al |q_{3x}|\right) \Theta\left(\eta_0+
\ln\frac{\al Q^2_{13}\>e^{-y}}{Q_{12}\sqrt{\ka^2+m^2}}\right).
\end{split}
\end{equation}
Again, in the region of $k^*$ emitted in the $P^*_1$ hemisphere
($\al<\sqrt{\ka^2+m^2}/{Q_{13}}$) the rapidity cut $\eta_0$ gives the
lower limit of the $\al$-integral.  In the other region of $k^*$ emitted
in the $P^*_3$ hemisphere ($\al>\sqrt{\ka^2+m^2}/{Q_{13}}$) it is the
recoil component $q_{3x}$ which provides the upper limit of $\al$. We
have
\begin{equation}
\al_m<\al<\al_M\>,\qquad
\al_m\equiv
\frac{\sqrt{\ka^2+m^2}\,Q_{12}\,e^{y-\eta_0}}{Q^2_{13}}\>,
\quad
\al_M \equiv {\frac{\sqrt{\ka^2+m^2}}{|q_{3x}|}}\>,
\end{equation}
giving
\begin{equation}
  \label{eq:I13}
\begin{split}
I_{13}&=\int_{-\pi}^{\pi }\frac{d\phi}{2\pi} 
\int_{\al_m}^{\al_M}\frac{d\al}{\al}
\left(\left|\sqrt{\ka^2\!+\!m^2}\cos\phi\!+\!\al q_{3x}\right|
\!-\!\al |q_{3x}|\right) \\
&= \frac{2}{\pi}\sqrt{\ka^2+m^2}
\left(\eta_0-y+\ln\frac{Q_{13}}{Q_{12}}+
\ln\frac{Q_{13}\zeta}{|q_{3x}|}\right),\qquad \zeta=2e^{-2}\>. 
\end{split}
\end{equation}
Here $\zeta$ comes from the integration region of large rapidity of
$k$ near $p_3$. In conclusion we have
\begin{equation}
\begin{split}
\de r_{13}= \nu\cp\left(\eta_0-y+\ln\frac{Q_{13}}{Q_{12}}
+\ln\frac{Q_{13}\zeta}{|q_{3x}|}\right).
\end{split}
\end{equation}
A similar result is obtained from the last $23$-dipole radiator:
\begin{equation}
\de r_{23}=\nu\cp\left(\eta_0+y+\ln\frac{Q_{23}}{Q_{12}}
+\ln\frac{Q_{23}\zeta}{|q_{3x}|}\right)\>.
\end{equation}
A compilation of these three contributions then gives the result 
\eqref{eq:Rad-NP}.

\section{Distribution \label{App:Distribution}}
\subsection{Evaluation of $\cA_{\conf}^{\PT}$}
Here we compute $\cA^{\PT}_{\conf}$ obtained from \eqref{eq:Mellin}
by taking only the PT part of the radiator given in \eqref{eq:Rad-PT}.
Since $R^{\PT}_{\conf}$ does not depend on the recoil, in
\eqref{eq:hatsigma} we can freely integrate over $q_{3x}$ to get
\begin{equation}
\label{eq:sigmaPT}
\cA^{\PT}_{\conf}(\sh,Q_t,\Ko)\>=\>
\int\frac{d\nu\,e^{\nu\Ko}}{2\pi i\nu}\>\int_{-\infty}^{\infty}
\frac{d\be\,e^{-R^{\PT}_{\conf}\left(\bnu\sqrt{1+\be^2}\right)}}
{\pi(1+\be^2)}\>,
\qquad \bnu=e^{\gam_E}\nu\>.
\end{equation}

We now perform the Mellin transform to SL accuracy.  We make use of
the operator identity
\begin{equation}
\label{eq:op-identity}
\int\frac{d\nu\,e^{\nu\Ko}}{2\pi i\nu}\>G(\nu) = \frac{1}{\Gamma
\left(1+\frac{\partial}{\partial\ln\Ko}\right)}\>G(\Ko^{-1})
\end{equation}
for any logarithmically varying function $G$. (To prove this,
multiply both sides by the $\Gamma$-function operator and use
the definition
$\Gamma(z) = \int_0^\infty dx\,x^{z-1}e^{-x}$.)
Thus we obtain the quantity $\cA_\de^{\PT}$ in the form
\begin{equation}
\cA_\de^{\PT}=\int_{-\infty}^{\infty}\frac{d\be}{\pi(1+\be^2)}\>
\frac{1}{\Gamma\left(1+\frac{\partial}{\partial\ln\Ko}\right)}\>
e^{-R^{\PT}_{\conf}\left(\bKo^{-1}\sqrt{1+\be^2}\right)}\>,\quad
\bKo=e^{-\gam_E}\Ko\>.
\end{equation}
We make a logarithmic expansion of the radiator, neglecting 
contributions from the second logarithmic derivative, which are beyond 
SL accuracy. So we obtain
\begin{equation}
  \label{eq:RadSL}
\begin{split}
& R_{\conf}^{\PT}\left(\bKo^{-1}\sqrt{1+\be^2}\right) = 
R_{\conf}^{\PT}\left(\Ko^{-1}\right)
+\left(\gamma_E+\ln\sqrt{1+\be^2}\right)R'\>,\\
& R' = -\frac{\partial}{\partial\ln\Ko}
R_\conf^{\PT}\left(\Ko^{-1}\right)\>.
\end{split}
\end{equation}
To SL accuracy, $R'$ is given by \eqref{eq:cF}.

Therefore we may write
\begin{equation}
\cA_\de^{\PT}=e^{-R^{\PT}_{\conf}(\Ko^{-1})}
\frac{e^{-\gamma_E R'}}{\Gamma(1+R')}\>
\int_{-\infty}^{\infty}\frac{d\be}{\pi(1+\be^2)^{1+\frac{1}{2}R'}}\>,
\end{equation}
which leads immediately to the result \eqref{eq:sigPT}.

\subsection{Including the NP correction \label{App:NPdist}}
The analysis is similar to the one in \cite{Tmin}. We report only the
essential steps. 
Consider 
\begin{equation}
  \label{eq:sigmahat}
\begin{split}
  \hat \cA_\conf(\nu) &=
\int_{-\infty}^{\infty}\frac{\nu d\be dq_{3x}}{2\pi}
e^{-\nu(|q_{3x}| - i\be q_{3x})}
e^{-R^{\PT}_{\conf}\left(\bnu\sqrt{1+\be^2}\right)}\\
&\qquad\times\left\{1-\nu\,\cp\,\left(C_1^{(\conf)}(\eta_0-\eta_3)+
C_2^{(\conf)}(\eta_0+\eta_3)+
C_3^{(\conf)}\ln\frac{\zeta Q_t}{|q_{3x}|}\right)\right\},
\end{split}
\end{equation}
where we have expanded $\exp(-\de R_{\conf})$ to first order in order
to obtain the leading correction. Performing the $q_{3x}$ integration
we get
\begin{equation}
  \label{eq:sigmahat1}
  \hat \cA_{\conf}(\nu)=  \hat \cA_{\conf}^{\PT}(\nu)-
\nu\,\cp\>f_{\conf}(\nu)\>,
\end{equation}
where
\begin{equation}
  \label{eq:f}
f_{\conf}(\nu)\!=\!\!\int_{-\infty}^{\infty}\!\!
\frac{d\be\,e^{-R^{\PT}_{\conf}\left(\bnu\sqrt{1+\be^2}\right)}}
{\pi(1+\be^2)}
\left(C_1^{(\conf)}(\eta_0\!-\!\eta_3)+C_2^{(\conf)}(\eta_0\!+\!\eta_3)+
C_3^{(\conf)}[\ln\zeta\bnu Q_t\!+\!\chi(\be)]\right), 
\end{equation}
with $\chi(\be)=\ln\sqrt{1+\be^2}+\be\tan^{-1}\be$.
This gives
\begin{equation}
  \label{eq:desigma}
  \cA_{\conf}(\Ko)=\cA_{\conf}^{\PT}(\Ko)+\de\cA_{\conf}(\Ko)\>,
\quad
  \de\cA_{\conf}(\Ko)=-\cp \partial_{\Ko}
\int\frac{d\nu e^{\nu\Ko}}{2\pi i \nu}\,f_{\conf}(\nu)\>.
\end{equation}
From equation \eqref{eq:op-identity} and \eqref{eq:desigma} we obtain
\begin{equation}
\begin{split}
&\de\cA_\conf(\Ko)=\frac{-\cp}{\Ko}\int_{-\infty}^{\infty}
\frac{d\be}{\pi(1+\be^2)}\frac{\frac{\partial}{\partial\ln\Ko}}
{\Gamma\left(1+\frac{\partial}{\partial\ln\Ko}\right)}\\
&\times e^{-R^{\PT}_{\conf}\left(\bKo^{-1}\sqrt{1+\be^2}\right)}
\left(C_1(\eta_0-\eta_3)+C_2(\eta_0+\eta_3)+
C_3[\ln(\zeta\bKo^{-1} Q_t)+\chi(\be)]\right)\>.
\end{split}
\end{equation}
Neglecting contributions from the second logarithmic derivative 
of $R^{\PT}_{\conf}$, which are beyond SL accuracy, we obtain
\begin{equation}
\begin{split}
&\de\cA_\conf(\Ko)=\frac{-\cp}{\Ko}e^{-R^{\PT}_{\conf}(\Ko^{-1})}
\frac{R' e^{-\gamma_E R'}}{\Gamma(1+R')}
\int_{-\infty}^{\infty}
\frac{d\be}{\pi(1+\be^2)^{1+\frac{1}{2}R'}}\\
&\quad\times\left[C_1^{(\conf)}(\eta_0\!-\!\eta_3)
  +C_2^{(\conf)}(\eta_0\!+\!\eta_3)+
  C_3^{(\conf)}\left\{\ln(\zeta\bKo^{-1} Q_t)+\chi(\be)+
\psi(1+R')-\frac{1}{R'}\right\}\right]\,.
\end{split}
\end{equation}
Performing the $\be$ integral then gives
\begin{equation}
\de\cA_\conf(\Ko)=\frac{-\cp}{\Ko}\Delta_\conf(\Ko)
R'(\Ko)\cA_\conf^{\PT}(\Ko)
=-\cp\Delta_\conf(\Ko)\partial_{\Ko}\cA_\conf^{\PT}\>,
\end{equation}
in other words the distribution is shifted by $-\cp\Delta_\conf$,
where $\Delta_\conf$ is given in \eqref{eq:Delta}.

\section{Matching\label{App:matching}}
Expanding the numerator of the integrand of \eqref{eq:SigPT-fine} we
have (flavour and configuration indices are understood)
\begin{equation}
  \label{eq:cpi}
\begin{split}
& C(\as)\cdot\cP(X_1,X_2,\Ko)\cdot \cA(\sh,Q_t,\Ko)=\cP(X_1,X_2,K)
\\&
\cdot \left\{1+\frac{\as}{2\pi}\left(-2C_TL^2+\bar{G}_{11}\,L+
c_1+\bar{r}_1\right)
+\cO{\as^2}\right\}\>, \quad L=\ln\frac{M}{\Ko}\>,
\end{split}
\end{equation}
where we have expanded around $\Ko=M$. Here $c_1$ is the first order
term of the coefficient function $C(\as)$ in \eqref{eq:Cexp} and we
have
\begin{equation}
  \label{eq:GC}
\begin{split}
\bar{G}_{11} &= \gamma - 4\sum_a C_a\ln\frac{Q_a}{M}\>, \\
\bar{r}_1 &= \gamma\ln\frac{K}{M} - 
2\sum_a C_a\left(\ln^2\frac{Q_a}{M}-\ln^2\frac{Q_a}{K}\right).
\end{split}
\end{equation}
The quantity $\gamma(X_1,X_2,K)$ is defined by
\begin{equation}
  \label{eq:gamma}
\frac{\as}{2\pi}\cP(X_1,X_2,K) \cdot \gamma = 
-\frac{\partial\cP(X_1,X_2,K)}{\partial\ln K}
\end{equation}
Therefore the first order contribution $\Sigma_{1}$ in \eqref{eq:Sig1}
is:
\begin{equation}
  \label{eq:Sig_logs}
\Sigma_1 = -2C_T\,L^2+G_{11}\,L+C_1\>, \qquad C_1=c_1+r_1\>,
\end{equation}
with 
\begin{equation}
\label{eq:G11}
G_{11}(Q_m)= \VEV{\bar{G}_{11}}\>, \qquad r_1=\VEV{\bar{r}_1}\>,
\end{equation}
where we introduced the averages
\begin{equation}
  \label{eq:average}
\VEV{F}=\sigma^{-1}(Q_m)
\int_{Q_m}^{Q_M}\!\!dQ_t 
\int dX_1dX_2\sum_{f\conf}\left\{\frac{d\hat \sigma^{f}_{\conf}}{dQ_t}
\, \cP^{f}_{\conf}(X_1,X_2,K)\cdot F_{\conf}\right\}\>.
\end{equation}
The normalization $\sigma(Q_m)$ is given by \eqref{eq:sigQm}.  Notice
that $r_1=0$ for $K=M$.

\section{Particle production from beam remnant}
\label{App:beam}
Here we discuss a factorized model for the beam remnant interaction
which gives the contribution $\Ko^{\remnant}$ to the shift as in
\eqref{eq:Krem}.  We assume an independent emission model, roughly
similar to the one discussed in \cite{MW,MW88}, in which the two
outgoing hadron remnants produce soft particles with distribution in
$\vec{k}_t$ and rapidity given by
\begin{equation}
  \label{eq:dnsoft}
\frac{dn^{\remnant}}{d^2k_td\eta}=\frac{\cN b^2}{2\pi} e^{-bk_t}\>,
\qquad \VEV{k_t}^{\remnant}=\frac{2}{b}\>,
\end{equation}
where $\cN$ is the number per unit rapidity of soft hadrons emitted by
the beam remnants and $\VEV{k_t}^{\remnant}$ the $k_t$ mean value.
Then we obtain an additional term in the radiator due to the beam
remnant
\begin{equation}
  \label{eq:Radsoft}
\begin{split}
R^{\remnant} &= \frac{\cN b^2}{2\pi}\int d^2k_te^{-bk_t}
\left(1-e^{-\nu k_t|\cos\phi|}\right)\int_{-\eta_0}^{\eta_0}d\eta \\
&=2\eta_0\,\cN\left\{1-\frac{2b}{\pi(b^2-\nu^2)}\left(
\frac{2b^2}{\sqrt{b^2-\nu^2}}\tan^{-1}\sqrt{\frac{b-\nu}{b+\nu}}-\nu\right)
\right\}.
\end{split}
\end{equation}
For $\nu\sim b$, i.e. $\Ko\sim \VEV{k_t}^{\remnant}$ we get
distortions of the distribution.  For $\nu\ll b$, i.e. $\Ko\gg
\VEV{k_t}^{\remnant}$, we may expand
\begin{equation}
R^{\remnant} \simeq \nu\,\frac{8\eta_0 \cN }{\pi}\,b^{-1} + \ldots
= \nu\,\Ko^{\remnant} +\ldots 
\end{equation}
with $\Ko^{\remnant}$ given by \eqref{eq:Krem}. We conclude that,
in the region we consider, the factorized beam remnant interaction can
be taken into account simply as ``rigid shift''
\begin{equation}
  \label{eq:shiftrem}
\Ko\to\Ko-\Ko^{\remnant}\>,
\end{equation}
of the hard QCD distribution.


\begin{thebibliography}{50}
\bibitem{PTstandard}
% Thrust in E+E-
S.~Catani, L.~Trentadue, G.~Turnock and B.R.~Webber,
  \npb{407}{1993}{3}; \\
% BROADENING IN E+ E- ANNIHILATION.
S.~Catani, G.~Turnock and B.R.~Webber, \plb{295}{1992}{269};\\
% C-PARAMETER IN E+ E- ANNIHILATION.
S. Catani and B.R. Webber, \plb{427}{1998}{377}
[hep-ph/9801350];\\
% JET BROADENING MEASURES IN E+ E- ANNIHILATION.
Yu.L. Dokshitzer, A. Lucenti, G. Marchesini and G.P. Salam,
\jhep{01}{1998}{011} [hep-ph/9801324].

\bibitem{Tmin}
A.~Banfi, Yu.L.~Dokshitzer, G.~Marchesini and  G.~Zanderighi, 
\jhep{07}{2000}{002} [hep-ph/0004027];
\plb{508}{2001}{269} [hep-ph/0010267] and 
\jhep{03}{2001}{007} [hep-ph/0101205]. 

\bibitem{Dpar}
A.~Banfi, Yu.L.~Dokshitzer, G.~Marchesini and G.~Zanderighi, 
\jhep{05}{2001}{040} [hep-ph/0104162].

\bibitem{NPstandard}
B.R.~Webber, \plb {339}{1994}{148} [hep-ph/9408222]; 
see also {\em Proc. Summer School on Hadronic Aspects
of Collider Physics}, Zuoz, Switzerland, August 1994, 
ed. M.P. Locher (PSI, Villigen, 1994) [hep-ph/9411384]; \\
% gluon mass, non-analyticity, thrust
M.~Beneke and V.M.~Braun, \npb {454}{1995}{253} [hep-ph/9506452];\\
Yu.L.~Dokshitzer and B.R.~Webber,
\plb {352}{1995}{451} [hep-ph/9504219]; \\
%LEADING POWER CORRECTIONS IN QCD: FROM RENORMALONS TO PHENOMENOLOGY
R.~Akhoury and V.I.~Zakharov, \plb{357}{1995}{646}
[hep-ph/9504248];
\npb {465}{1996}{295} [hep-ph/9507253]; \\
%NONPERTURBATIVE CORRECTIONS IN RESUMMED CROSS-SECTIONS 
G.P. Korchemsky and G. Sterman,
\npb {437}{1995}{415}  [hep-ph/9411211]; \\
Yu.L.~Dokshitzer, V.A.~Khoze and S.I.~Troyan,
  \prd{53}{1996}{89} \mbox{[hep-ph/9506425]}; \\
P.~Nason and B.R.~Webber, \plb{395}{1997}{355}
[hep-ph/9612353]; \\
P.~Nason and M.H.~Seymour, \npb {454}{1995}{291}
[hep-ph/9506317]; \\
%EEC
Yu.L.~Dokshitzer, G.~Marchesini and B.R.~Webber, \jhep {07}{1999}{012}
\mbox{[hep-ph/9905339]}; \\
%Renormalons
 M.~Beneke, \prep {317}{1999}{1} \mbox{[hep-ph/9807443]}; \\
%DISENTANGLING RUNNING COUPLING AND CONFORMAL EFFECTS IN QCD.
S.J.~Brodsky, E.~Gardi, G.~Grunberg, J.~Rathsman,
\prd{63}{2001}{094017} [hep-ph/0002065];\\
%Renormalon resummation and exponentiation of soft and collinear gluon 
%radiation in the thrust distribution
E.~Gardi and J.~Rathsman, [hep-ph/0103217].

\bibitem{DMW}
Yu.L. Dokshitzer, G.\ Marchesini and  B.R.\ Webber,
\npb{469}{1996}{93} \mbox{[hep-ph/9512336].}

\bibitem{Milan}
Yu.L.~Dokshitzer, A.~Lucenti, G.~Marchesini and  G.P.~Salam, 
\npb{511}{1998}{396}, \mbox{[hep-ph/9707532]}, erratum
  \ibid{B593}{2001}{729}; \jhep{05}{1998}{003} [hep-ph/9802381];\\
M. Dasgupta and B.R. Webber \jhep{10}{1998}{001} [hep-ph/9809247].

\bibitem{Milan2}
 M.~Dasgupta, L.~Magnea and G.~Smye,
\jhep{11}{1999}{25} [hep-ph/9911316];\\
G.~Smye, \jhep{05}{2001}{005} [hep-ph/0101323]. 

\bibitem{broad}
Yu.L. Dokshitzer, G.~Marchesini and G.P.~Salam,
\epjc{3}{1999}{1} \mbox{[hep-ph/9812487]}.

\bibitem{Exp-shape}
% comparison theory-data for shapeV
P.~A.~Movilla Fernandez, O.~Biebel, S.~Bethke,
%``Tests of power corrections to event shape distributions from e+e-
%annihilation''
paper contributed to the EPS-HEP99 conference in Tampere, Finland,
hep-ex/9906033;\\
H.~Stenzel,
%``Power corrections to e+ e- event shape variables,''
MPI-PHE-99-09 {\it Prepared for 34th Rencontres de Moriond:} 
``QCD and Hadronic interactions'', Les Arcs, France, 20-27 Mar 1999;\\
%
ALEPH Collaboration,
%``QCD Measurements in e+e- Annihilations at Centre-of-Mass 
%Energies between 189 and 202 GeV'',
ALEPH 2000-044 CONF 2000-027;\\
P.~Abreu {\it et al.}  (DELPHI Collaboration),
%``Energy dependence of event shapes and of alpha(s) at LEP-2,''
\plb{456}{1999}{322};\\
%
DELPHI Collaboration, 
%``The Running of the Strong Coupling and a Study
%of Power Corrections to Hadronic Event Shapes with the DELPHI Detector
%at LEP'', 
DELPHI 2000-116 CONF 415, July 2000;\\
%
M.~Acciarri {\it et al.}  (L3 Collaboration),
%``QCD studies in e+ e- annihilation from 30-GeV to 189-GeV,''
\plb{489}{2000}{65} [hep-ex/0005045].

\bibitem{Exp-running}
% running as measurements
D.~Decamp {\it et al.}  (ALEPH Collaboration),
%``Measurement of alpha-s in hadronic Z decays using all orders resummed
%predictions,''
\plb{284}{1992}{163};\\
P.~Abreu, {\it et al.}  (DELPHI Collaboration)
%``Consistent measurements of \as from precise oriented event shape
%distributions''
\epjc{14}{2000}{557} [hep-ex/0002026];\\
P.~A.~Movilla Fernandez, O.~Biebel, S.~Bethke, S.~Kluth and
P.~Pfeifenschneider   (JADE Collaboration),
%``A study of event shapes and determinations of alpha(s) using data of
%e+ e- annihilations at s**(1/2) = 22-GeV to 44-GeV,''
\epjc{1}{1998}{461} [hep-ex/9708034];\\
M.~Acciarri {\it et al.}  (L3 Collaboration),
%``Study of hadronic events and measurements of alpha(s) between 30-GeV
%and 91-GeV,''
\plb{411}{1997}{339};\\
P.~D.~Acton {\it et al.}  (OPAL Collaboration),
%``A Determination of alpha-s (M (Z0)) at LEP using resummed QCD
%calculations,''
\zpc{59}{1993}{1};\\
K.~Abe {\it et al.}  (SLD Collaboration),
%``Measurement of alpha-s (M(Z)**2) from hadronic event observables at the
%Z0 resonance,''
\prd{51}{1995}{962} [hep-ex/9501003].

\bibitem{DDTetc} 
  
  Yu. L. Dokshitzer, D.I. Dyakonov and S.I. Troyan,  \prep{58}{1980}{270};\\
  A. Bassetto, M. Ciafaloni and G. Marchesini, \prep{100}{1983}{201};\\
  for recent applications see
  A. Guffanti and G. Smye, \jhep{10}{2000}{025} [hep-ph/0007190],
  and, in DIS thrust distribution, V.~Antonelli, M.~Dasgupta and G.P.~Salam,
  \jhep{02}{2000}{001} [hep-ph/9912488].
  
\bibitem{BG} S.J.~Burby and E.W.N.~Glover, \jhep{04}{2001}{029}
  [hep-ph/0101226].

\bibitem{DS}
M.~Dasgupta and G.P.~Salam, hep-ph/0104277.

\bibitem{MW}
G.~Marchesini and B.R.~Webber, \prd{38}{1988}{3419}.  

\bibitem{U1}UA1 Collaboration, C. Albajar et al. \npb{309}{1988}{405}.
  
\bibitem{MCworkshops} R.K.~Ellis et al., ``Report of the QCD Tools
  Working Group'',  hep-ph/0011122.

\bibitem{MW88} G.~Marchesini and B.R.~Webber, \npb{310}{1988}{461}.

\bibitem{CMW}
S.~Catani, G.~Marchesini and B.R.~Webber,  
\npb{349}{1991}{635}.

\bibitem{DYRAD}
W.T.~Giele, E.W.N.~Glover, D.A.~Kosower, \npb{403}{1993}{633}
[hep-ph/9302225]. 

\bibitem{GZ}
G.P.~Salam and G.~Zanderighi, 
\npps{86}{2000}{430} [hep-ph/9909324].

\bibitem{SF}
%"Parton Distributions and the LHC: W and Z Production", 
A.D.~Martin, R.G.~Roberts, W.J.~Stirling and R.S.~Thorne, 
%Univ. Durham preprint DTP/99/64 (1999), 
\epjc{14}{2000}{133} [hep-ph/9907231].

\bibitem{KS}
G.P.~Korchemsky and  G.~Sterman, \npb{555}{1999}{335} [hep-ph/9902341];\\
G.P.~Korchemsky and  S.~Tafat, \jhep {10}{2000}{010} [hep-ph/0007005].

\end{thebibliography}
\end{document}